\documentclass[]{aa} 

\usepackage{graphicx}
\usepackage{txfonts}
\usepackage{natbib}
\usepackage{longtable}
\usepackage{lscape}
\usepackage{color}

\providecommand{\sorthelp}[1]{}

\newcommand{\Herschel}{{\it Herschel }}

\begin{document}
 
\title{
{Synthetic observations of dust emission and polarisation of Galactic cold clumps}
}

\author{Mika  Juvela\inst{1},
Paolo Padoan\inst{2}
Isabelle      Ristorcelli\inst{3,4},
Veli-Matti Pelkonen\inst{5}
}

\institute{
Department of Physics, P.O.Box 64, FI-00014, University of Helsinki,
Finland, {\em mika.juvela@helsinki.fi}
\and
ICREA \& Institut de Ci\`encies del Cosmos, Universitat de Barcelona,
IEEC-UB, Mart\'i Franqu\`es 1, E08028 Barcelona
\and 
Universit\'e de Toulouse, UPS-OMP, IRAP, F-31028 Toulouse cedex 4, France   
\and 
CNRS, IRAP, 9 Av. colonel Roche, BP 44346, F-31028 Toulouse cedex 4, France  
\and
Institut de Ci\`encies del Cosmos, Universitat de Barcelona,
IEEC-UB, Mart\'i Franqu\`es 1, E08028 Barcelona
}

\authorrunning{M. Juvela et al.}

\date{Received September 15, 1996; accepted March 16, 1997}

\abstract { 
The Planck Catalogue of Galactic Cold Clumps (PGCC) contains over 13000
sources that are detected based on their cold dust signature. They
are believed to consist of a mixture of quiescent, pre-stellar, and
already star-forming objects within molecular clouds.
} 
{
We extracted PGCC-type objects from cloud simulations and examined their
physical and polarisation properties. The comparison with the PGCC
catalogue helps to characterise the properties of this large sample of
Galactic objects and, conversely, provides valuable tests for numerical
simulations of large volumes of the interstellar medium and the
evolution towards pre-stellar cores.
}
{
We used several magnetohydrodynamical (MHD) simulation snapshots to
define the density field of our model clouds. Sub-millimetre images of
the surface brightness and polarised signal were obtained with
radiative transfer calculations. We examined the statistics of 
synthetic cold clump catalogues extracted with methods similar to the
PGCC. We also examined the variations of the polarisation fraction $p$ in
the clumps.
}
{
The clump sizes, aspect ratios, and temperatures in the synthetic
catalogue are similar to the PGCC. The fluxes and column densities of
synthetic clumps are smaller by a factor of a few. Rather than with an
increased dust opacity, this could be explained by increasing the
average column density of the model by a factor of two to three, close
to $N({\rm H})= 10^{22}$\,cm$^{\-2}$. When the line of sight is
parallel to the mean magnetic field, the polarisation fraction tends
to increase towards the clump centres, which is contrary to
observations. When the field is perpendicular, the polarisation
fraction tends to decrease towards the clumps, but the drop in $p$ is
small (e.g. from $p\sim$8\% to $p\sim$7\%).
}
{ 
{Magnetic field geometry reduces the polarisation fraction in the
simulated clumps by only $\Delta p\sim$1\% on average. The larger drop
seen towards the actual PGCC clumps therefore suggests some loss of grain
alignment in the dense medium, such as predicted by the radiative torque
mechanism. The statistical study is not able to quantify dust opacity
changes at the scale of the PGCC clumps.}
}


\maketitle

\section{Introduction} \label{sect:intro}

The Planck Catalogue of Galactic Cold Clumps (PGCC) \citep{PGCC}
contains over 13000 cold Galactic sources that were detected using
Planck sub-millimetre observations \citep{planck2014-a01} and the
100\,$\mu$m data from the The Infrared Astronomical Satellite (IRAS)
survey \citep{Neugebauer1984}. The source extraction was not based on
the absolute brightness of the sources but on the cold dust signature,
that is obtained by subtracting the signal of warmer dust, that is
traced with 100\,$\mu$m surface brightness, from the Planck data
\citep{Montier2010}. The typical dust colour temperatures of PGCC
sources is $T_{\rm d}<14$\,K. In the interstellar medium (ISM), such
low temperatures are found only in regions of high column density
where the ISM is shielded from the interstellar radiation field by the
high optical depths caused by dust. Apart from minor contamination by
external galaxies, PGCC sources should therefore correspond to dense
regions of Galactic molecular clouds where star formation may take
place if it is not already.

The PGCC covers sources with distances from $d\sim$100\,pc up to
several kiloparsecs. This combined with the $\sim 4.5\arcmin$ full
width at half maximum (FWHM) beam size of the Planck and IRAS
observations, for the analysis further convolved to 5$\arcmin$, means
that the catalogue contains a heterogeneous sample of objects from
gravitationally bound cloud cores in nearby clouds to entire clouds at
kiloparsec distances. We use the term clump to refer to both the
observed and simulated PGCC sources, irrespective of their physical
size. The detection algorithm and its parameters have their own impact
on the contents of the catalogue; they set preference on sources that
are close to the beam size and have the largest temperature contrast
relative to their immediate environment (as seen in projection via the
colour temperature). This also contributes to the fact that PGCC
contains different types of sources at different distances.

It is important to note that PGCC sources have been the target of many
follow-up observations that mapped their molecular line emission
\citep{Wu2012, Meng2013, Parikka2015, Zhang2016, Liu2016ApJS222_7,
Feher2017, Zhang2018} and looked at their internal structure with
higher-resolution continuum observations with \Herschel, for example, 
\citep{PlanckII, GCC-III, GCC-IV, GCC-VII,GCC-IX} or the SCUBA-II
instrument at JCMT \citep{SCOPE, Juvela_2018_pilot}. These
observations show multiple levels of fragmentation below the scales
resolved by Planck and, in spite of the low temperatures, many PGCC
clumps are already actively forming stars. The range of evolutionary
phases is also reflected in the chemical properties
\citep{Tatematsu2017} and dust grain properties \citep{GCC-V, GCC-VI}.

Our knowledge of magnetic fields in molecular clouds is based mainly
on light polarisation, the optical and near-infrared (NIR)
observations of background stars
\citep{Goodman1995,Whittet2001,Pereyra2004,Alves2008,Chapman2011,Cox2016,Neha2018,Kandori2018},
and the polarised dust emission at far-infrared (FIR), sub-millimetre,
and radio wavelengths \citep{WardThompson2000, Koch2014, Matthews2014,
Fissel2016, Pattle2017_BISTRO}. The Planck survey provides a large
amount of data for polarisation studies at cloud scales
\citep{Planck_2015_XX,Planck_2015_XIX,Planck2016_XXXIII}. The Planck
data have been used especially to study the polarisation fraction and
the correlations in the relative morphology of column density and
magnetic field structures
\citep{Planck_2015_XX,Planck2016_XXXIII,Malinen2016,Soler2016,Alina2019}.
Particularly, the drop of polarisation fraction $p$ towards PGCC
clumps has been observed with high significance in the Planck 353\,GHz
data \citep{Alina2019, Ristorcelli2019}. The variations of $p$ are
interesting because they are related to the configuration of the
magnetic fields in clumps and cores during the star formation process.
However, $p$ is also affected by variations in the efficiency of the
grain alignment, as predicted, for example, by the theory of radiative
torque alignment (RAT) \citep{Lazarian1997, ChoLazarian2005,
HoangLazarian2014} and demonstrated by numerical simulations
\citep{Pelkonen2009, Brauer2016, Reissl_2018_filaments}. These suggest
that high optical depths and more frequent gas collisions should
significantly reduce the grain alignment and thus the polarised
emission observable from within the clumps. The PGCC provides a
statistically significant sample to study these questions
observationally, although the Planck resolution limits the
investigations to structures that are typically much larger than an
individual cloud core. However, polarisation of selected PGCCs has
already been studied at higher resolution with the SCUBA-2 POL-2
instrument at JCMT \citep{Liu_G35_pol,Juvela2018_POL,Liu2018_G9.62}, and many
more will be covered by ongoing surveys \citep{BISTRO}.

In this paper we compare synthetic observations of PGCC-type objects
to the sources in the Planck catalogue. We use several snapshots of
MHD simulations of supernova-driven turbulence that provide a large
sample of dense clumps and cores. Radiative transfer calculations are
used to produce synthetic observations in the Planck and IRAS bands.
We extract from these maps sources (clumps) with an algorithm that
closely follows the procedures used in the creation of the PGCC
catalogue. Radiative transfer calculations also provide predictions
for polarised intensity that will be obtained under the assumption of
constant grain alignment. With these data, we can examine the
polarisation fraction variations (geometrical depolarisation) that are
caused by the magnetic field geometry alone. In future studies, these
will be compared to observations to assess the importance of grain
alignment variations. The comparison of simulations and observations
helps us to better understand the physical nature of the PGCC objects.
Conversely, it also serves as a valuable test for the numerical
simulations, especially regarding the formation of dense structures as
precursors of star formation.

The content of the paper is as follows: In Sect.~\ref{met} we describe
the methods related to the MHD calculations (Sect.~\ref{met:MHD}), the
radiative transfer modelling (Sect.~\ref{met:RT}), and the creation of
the synthetic source catalogue (Sect.\ref{met:PGCC}). 
The results are presented in Sect.~\ref{res}, where we compare the
synthetic source catalogue with the actual PGCC catalogue
(Sect.~\ref{res:PGCC}), and make predictions for the polarisation
fraction in the clumps (Sect.~\ref{res:p}). We discuss the results in
Sect.~\ref{sect:discussion} before summarising our conclusions in
Sect.~\ref{sect:conclusions}.


\section{Methods}  \label{met}

In this section, we describe the MHD simulations (Sect.~\ref{met:MHD})
and the radiative transfer calculations (Sect.~\ref{met:RT}) that
provided synthetic surface brightness maps. Section~\ref{met:PGCC}
describes how cold clumps were then extracted from these observations.

\subsection{MHD simulations}  \label{met:MHD}

We used MHD simulations that are described in \citet{Padoan2016_SN-I}
and have been used, for example, for synthetic line observations of
molecular clouds \citep{Padoan2016_SN-III} and for studies of the
star-formation rate \citep{Padoan2017_SN-IV}. The simulations of
supernova-driven turbulence were run with the Ramses code
\citep{Teyssier2002} using a 250-pc box with periodic boundary
conditions. The runs started with zero velocity, a uniform density
$n{\rm (H)}$=5\,cm$^{-3}$, and a uniform magnetic field of
4.6\,$\mu$G. The self-gravity was turned on after 45\,Myr and the
simulations were then run for another 11\,Myr. We use 18 snapshots
covering this later time interval. The volume is fully sampled by a
512$^3$ cell regular grid and the MHD runs have a maximum of six
levels of refinement in the octree grids covering the densest regions.
Thus, the largest cell size is 0.49\,pc and the minimum cell size
$7.6\times 10^{-3}$\,pc.

The intent of the numerical experiment was to represent a generic
volume within a Galactic spiral arm. Accordingly, the mean column
density of the simulation is $30 M_{\odot} {\rm pc}^{-2}$, comparable
to that of the Perseus arm \citep{Heyer_Terebey_1998}, giving a total
mass of $1.9\times 10^6 M_{\odot}$. The supernova rate of $6.25$
Myr$^{-1}$ is somewhat conservative compared to the nearly three times
larger value derived from a standard Kennicutt-Schmidt relation
\citep{Kennicutt_1998} and the column density of the simulation, but
within the total observational scatter of that relation.

\subsection{Calculation of surface brightness maps} \label{met:RT}

The MHD runs provided the density field that is one of the inputs of
radiative transfer modelling. The radiative transfer runs used only the
512$^3$ root grid plus four levels of refinement, resulting in a
spatial resolution of 0.031\,pc in dense regions. Because the model
clouds are assumed to be at distances $d \le 100$\,pc, this
corresponds to an angular resolution of 1.05$\arcmin$ or better. This
is sufficient because, like in the case of the real PGCC catalogue,
the beam size of the synthetic observations is $5.0\arcmin$.

The models are illuminated by an external radiation field with
intensities consistent with the local solar neighbourhood
\citep{Mathis1983}. The average column density through the model
volume is $N({\rm H})=3.8\times 10^{21}$\,cm$^{-2}$, which corresponds
to a visual extinction of $A_{\rm V}\sim 1$\,mag. The average optical
depth to the box centre is about half of this value. The effective
optical depth, which determines the radiation field intensity, is
still smaller because it is dominated by the lowest $A_{\rm V}$
sightlines between a cell and the model boundary. As a result, the
radiation field intensity is relatively constant at large scales and
the most significant variations of radiation field (and of dust
temperature) are dominated by smaller-scale optically thick
structures. 

The dust properties were taken from \citet{Compiegne2011}. This dust
model is fitted against diffuse medium observations and may not be
representative of dense cores where the sub-millimetre dust
emissivity is expected to be higher \citep{Ossenkopf1994, Martin2011,
GCC-V}
However, our low 5$\arcmin$ resolution dilutes the signal, especially
at larger distances. Models with higher sub-millimetre dust opacity
will be considered in Sect.~\ref{sect:mod}. 

The radiative transfer problem was solved with the Monte Carlo
programme SOC \citep{Juvela2019_SOC}. The dust grains were assumed to
remain at equilibrium with the local radiation field, although the
IRAS $\lambda=100$\,$\mu$m band may have some contribution from
stochastically heated small grains. In the current study, the source
detection and the flux estimates are based on data where the warm
emission component is subtracted. To the first order, this will
eliminate the effects of diffuse small-grain emission. 

SOC calculations used a grid of 52 logarithmically spaced frequencies
between $1 \times 10^{11}$\,Hz and $1 \times 3\times 10^{15}$\,Hz to
describe the radiation field intensity in the model clouds.  Dust
temperatures and emission from each cell of the model clouds were
solved based on this information. Direct line-of-sight (LOS)
integration of the radiative transfer equation resulted in surface
brightness maps where the pixel size corresponded to the smallest cell
size of the model. The maps were calculated towards the three
coordinate axis directions. Each map thus covers a square region of
250\,pc$\times$250\,pc with 8192$\times$8192 pixels with the size of
0.0305\,pc$\times$0.0305\,pc. Maps were calculated for the
monochromatic wavelengths of 100\,$\mu$m, 350\,$\mu$m, 550\,$\mu$m,
and 850\,$\mu$m, which correspond to the IRAS 100\,$\mu$m band and the
857\,GHz, 545\,GHz, and 353\,GHz Planck bands.

Synthetic observations were calculated for 12 cloud distances that
were spaced logarithmically from 100\,pc to 10000\,kpc.  We added
noise to the maps, corresponding (in the final maps after convolution)
to 0.06\,MJy\,sr$^{-1}$, 0.01\,MJy\,sr$^{-1}$, 0.01\,MJy\,sr$^{-1}$,
and 0.001\,MJy\,sr$^{-1}$ at 100\,$\mu$m, 350\,$\mu$m, 350\,$\mu$m,
and 850\,$\mu$m, respectively \citep{Miville2005, Planck-2015-X}. We
also added a relative-noise component, 1\% for 100\,$\mu$m and 0.5\%
at the other wavelengths, which actually dominates the total errors
budget. The maps were convolved with a Gaussian beam to the final
5.0$\arcmin$ resolution and resampled onto 1$\arcmin$ pixels. With 18
snapshots, 12 distances, and three view directions, the total number
of maps is 648 per frequency.  However, the distance $d$=10000\,pc
rarely resulted in any clump detection and effectively the largest
distance is $d$=6600\,pc.

SOC was also used to calculate predictions of the 353\,GHz polarised
dust emission in the form of Stokes ($I$, $Q$, $U$) maps. This was
done assuming a constant grain alignment efficiency throughout the
model volume and a theoretical maximum polarisation fraction of
$p$=20\%. The polarisation fraction is defined as
\begin{equation}
p = \frac{\sqrt{ Q^2+U^2 }}{I}.
\end{equation}
Although we do not include noise in the polarisation simulations, we
used the modified asymptotic estimator of \citet{P14},
\begin{equation}
p_{\rm mas} = p - b^2 \frac{ 1-\exp(-p^2/b^2) }{2p},
\end{equation}
where $b^2$ is 
\begin{equation}
b^2 = \frac{
\sigma_U^{\prime 2} \cos^2(2\psi_0-\theta) + \sigma_Q^{\prime 2} (2\psi_0-\theta)
}{I_0^2} \, ,
\label{eq:b2}
\end{equation}
with 
\begin{equation}
\theta = \frac{1}{2} \mathrm{atan} \left(
\frac{2 \sigma_{QU}}{\sigma_Q^2-\sigma_U^2}
\right)\, ,
\end{equation}
\begin{equation}
\sigma_Q^{\prime 2} = 
\sigma_Q^2 \cos^2\theta + \sigma_U^2 \sin^2\theta
+ \sigma_{QU} \sin 2\theta,
\end{equation}
\begin{equation}
\sigma_U^{\prime 2} = 
\sigma_Q^2 \sin^2\theta + \sigma_U^2 \cos^2\theta
- \sigma_{QU} \sin 2\theta,
\end{equation}
where $\sigma_{\rm Q}$ and $\sigma_{\rm U}$ are the error estimates of
$Q$ and $U$ and $\sigma_{\rm QU}$ their covariance.
In Eq.~(\ref{eq:b2}) $\psi_0$ is the true polarisation angle that is
in practice replaced by its estimate 
\begin{equation}
\psi = 0.5 \arctan(U, Q).
\end{equation}

Apart from grain alignment efficiency, here assumed to be constant,
the polarisation fraction depends on the magnetic field geometry. For
a single LOS, the main factors are the angle $\gamma$ between the
plane-of-the-sky (POS) and the B-field direction and the variations of
the POS-projected magnetic field direction along the LOS. Because $Q$
and $U$ are proportional to $\cos ^2 \gamma$, we characterise the
first factor with the averaged quantity
\begin{equation}
\langle \cos^2 \gamma \rangle =
\frac{ \int R(\vec r) j_{\nu}(\vec r) \cos^2 \gamma(\vec r) dl }{
       \int R(\vec r) j_{\nu}(\vec r)                       dl },
\label{eq:gamma}       
\end{equation}
where $R(\vec r)$ is the polarisation reduction factor, $j_{\nu}$ the
dust emissivity at the observed wavelength, $l$ is distance, and the
integration extends over the full LOS \citep{Chen2016_B}. This
quantity is independent of the field geometry projected onto the POS
and, since $R$ is kept constant, is directly the emission-weighted
average of $\cos^2 \gamma$. The observed polarisation is largest when
the magnetic field is in the POS and thus when $\gamma$ is zero and
$\cos^2 \gamma$ is one. We denote with $\langle \gamma \rangle$ the
angle that corresponds to the $\langle \cos^2\gamma \rangle$ value
obtained from Eq.~(\ref{eq:gamma}).

The second factor is the variation of the POS-projected magnetic field
orientation along the LOS, which causes cancellation in the LOS $Q$
and $U$ integrals and thus leads to depolarisation. The effect can be
described using the polarisation angle dispersion function
\begin{equation}
S(\bar r)_{\rm LOS} = \sqrt{  
\frac{  
          \int j_{\nu, i}  \, (\Psi_i - \bar \Psi)^2  dl }{
          \int j_{\nu, i}  dl},
}
\label{eq:S_LOS}
\end{equation}
where the summation extends over all cells along the LOS, $\Psi$ is
the local polarisation angle (i.e. for a single cell), and $\bar \Psi$
is the similarly emission-weighted average angle. 

The quantity $S_{\rm LOS}$ only depends on data along a single LOS and
is thus different from the polarisation angle dispersion function
$S_{\rm POS}$ that can be derived from polarisation observations and
describes variations over the sky \citep{Planck_2015_XIX}. This can be
calculated as
\begin{equation}
S_{\rm POS}(\bar r, \delta) =
\sqrt{
\frac{1}{N}  \sum_{i=1}^N \left( \psi(\bar r) - \psi(\bar r + \bar
\delta_i) \right)^2
},
\label{eq:S_POS}
\end{equation}
where $\delta$ defines a spatial offset and the summation goes over
$N$ map pixels within distances [$\delta/2$, $3\delta/2$] from the
central position $\bar r$. We evaluate $S_{\rm LOS}$ from the
synthetic observations by setting $\delta$ equal to
$FWHM$/2=2.5$\arcmin$. The average $S_{\rm POS}$ value inside a
10$\arcmin$ radius circle is used to characterise the dispersion
associated to a clump.

\subsection{Clump catalogue} \label{met:PGCC}

The surface brightness maps at wavelengths
100\,$\mu$m\,--\,850\,$\mu$m were analysed to extract cold clumps
using a procedure that closely follows that of the PGCC study
\citep{PGCC}. In the following we describe the method in detail.

The 100\,$\mu$m map is compared in turn with each of the Planck maps.
At the location of each pixel, the average local colour of the dust
emission $C=I_{Planck}/I(100\mu{\rm m})$ is estimated as the median
over an annulus covering distances 5-15$\arcmin$ from the centre
position. A map of cold residual emission is calculated as Planck band
$I_{Planck}^{\rm CR} = I_{Planck} - C \times I(100\mu{\rm m})$, which
differs from zero only because of local variations in the dust SED.
Based on this map, the noise $\sigma^{\rm CR}$ of the cold residual is
estimated as the median absolute deviation of the pixel values within
an annulus from 5$\arcmin$ to 30$\arcmin$ from the centre. The
background level $I_{\rm bg}^{\rm CR}$ is estimated as the median over
the same pixels. Together these result in signal-to-noise (S/N) maps
of the cold residual, $SNR^{\rm CR}=(I_{Planck}^{\rm CR}-I_{\rm
bg}^{\rm CR})/\sigma^{\rm CR}$, one for each of the three Planck
bands. Source candidates are identified in these maps as local maxima
with $SNR^{\rm CR}>4$, further requiring that the value is the maximum
within a radius of 2$\arcmin$. The final merged catalogue contains
sources where each of the three bands has a source candidate and the
distances between the candidates are below 5$\arcmin$.

The second part of the clump analysis concerns the source fluxes. The
350\,$\mu$m (857\,GHz) source is fitted with a 2D Gaussian plus a
third order polynomial background model. The fit uses the position of
the detected source and returns estimates for its position angle and
FWHM sizes along its major and minor axis directions. These parameters
are used to separate a cold component in the 100\,$\mu$m emission. A
2D Gaussian fit is applied to the 100\,$\mu$m surface brightness map
using the previously fixed position and shape of the Gaussian. If
there are other sources within a radius of 10$\arcmin$, these are
fitted together as additional 2D Gaussian components (up to three
Gaussians). After the fits, a corrected 100\,$\mu$m warm template map
is obtained by subtracting the fitted Gaussians and this is used to
calculate new cold residual maps $I_{Planck}^{\rm CR}$ at the three
Planck wavelengths. The flux densities of the detected clumps are
estimated from these $I_{Planck}^{\rm CR}$ maps with aperture
photometry. In normalised distance units $r=x/{\rm FWHM}_x$, the
aperture extends to a distance of $r=2.0$. The fluxes are estimated
after subtracting the local background that is calculated as the
median over an annulus that extends over the distance range
$r=2.0-2.5$. If a source has a truly Gaussian shape, this results in
flux estimates that are some 70\% of the total intensity of the
Gaussian.

The remaining clump characteristics are calculated based on the
parameters derived above, based on synthetic observations, without
resorting to direct information about the density and temperature
values of the models. The source temperatures are estimated by fitting
the spectral energy distribution (SED) with a modified blackbody
function,
\begin{equation}
F_{\nu}  = F(\nu_0) \frac{B_{\nu}(\nu, T_{\rm d})}{B_{\nu}(\nu_0.
T_{\rm d})} 
\kappa(\nu_0) ( \nu/\nu_0)^{\beta},
\end{equation}
using the flux density $F(\nu_0)$ at a reference frequency $\nu_0$ and
the dust temperature $T_{\rm d}$ as free parameters. Following the
example of \citet{PGCC}, we use dust opacities
$\kappa_{\nu}=0.1(\nu/10^{12}\,{\rm Hz})^{\beta}$\,cm$^2$\,g$^{-1}$
\citep{Beckwith1990} and fix the spectral index to $\beta=2.0$. The
actual spectral index of the employed dust model is $\beta\sim1.84$
over the 100\,$\mu$m-850\,$\mu$m wavelength range. This means that the
SED fit will underestimate the dust temperature but only by a fraction
of one degree. Because distances $d$ are known for all of synthetic
clumps, their masses can be calculated as
\begin{equation}
M = \frac{F_{\nu} d^2}{B_{\nu}(T_{\rm d}) \kappa_{\nu}},
\end{equation}
where $B_{\nu}$ is the Planck function and $T_{\rm d}$ the colour
temperature obtained from the SED fit. The aperture size and the
measured flux density provide estimates of the average column density
of each clump. The fitted 100\,$\mu$m background component and the
values of the reference annuli used in the Planck photometry provide
the spectrum of the local warm background, which is used to estimate
the colour temperature and the column density of the warm background. 

Our procedures followed closely the methods used in \citet{PGCC}, with
only minor modifications. We fitted the cold residuals with a Gaussian
(or up to three Gaussians), as in the case of the PGCC catalogue.
However, the background was always modelled with a third order
polynomial while in \citet{PGCC} the degree ranged from three to six.
This can be justified by the smaller line-of-sight confusion of the
synthetic observations, especially when compared to Planck
observations of low Galactic latitudes.

\begin{figure*}
\includegraphics[width=18cm]{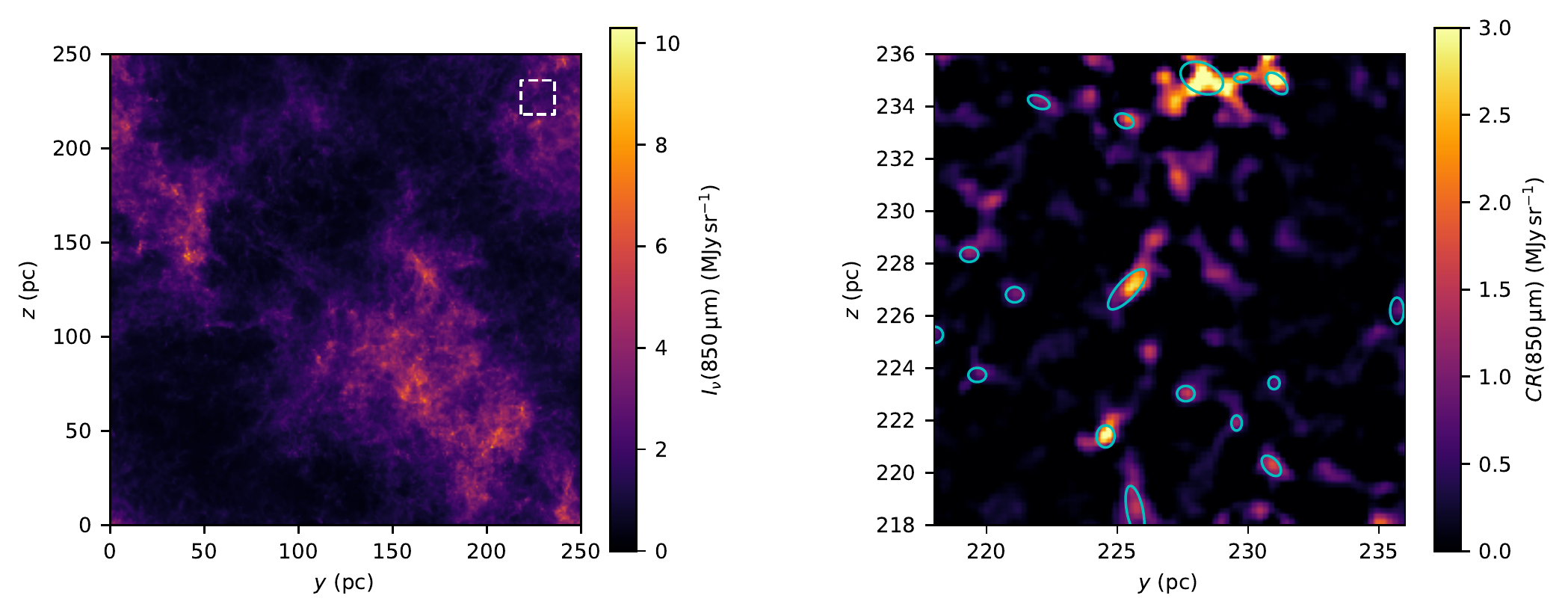}
\caption{
Example of surface brightness data and clump extraction. The left
frame shows the 850\,$\mu$m map of one snapshot (number 377), with the
view direction $x$ and assumed distance of $d=351\,$pc. The right
frame shows the cold residual at 350\,$\mu$m (857\,GHz) for the area
indicated with the dashed box in the first frame. The cyan ellipses
correspond to the clumps that have been detected with S/N above four at all
three Planck wavelengths and have reliable photometry in those bands
and at 100\,$\mu$m.
}
\label{fig:sample_maps}
\end{figure*}


\section{Results} \label{res}

\subsection{Clump catalogue} \label{res:PGCC}

\begin{figure}
\includegraphics[width=8.8cm]{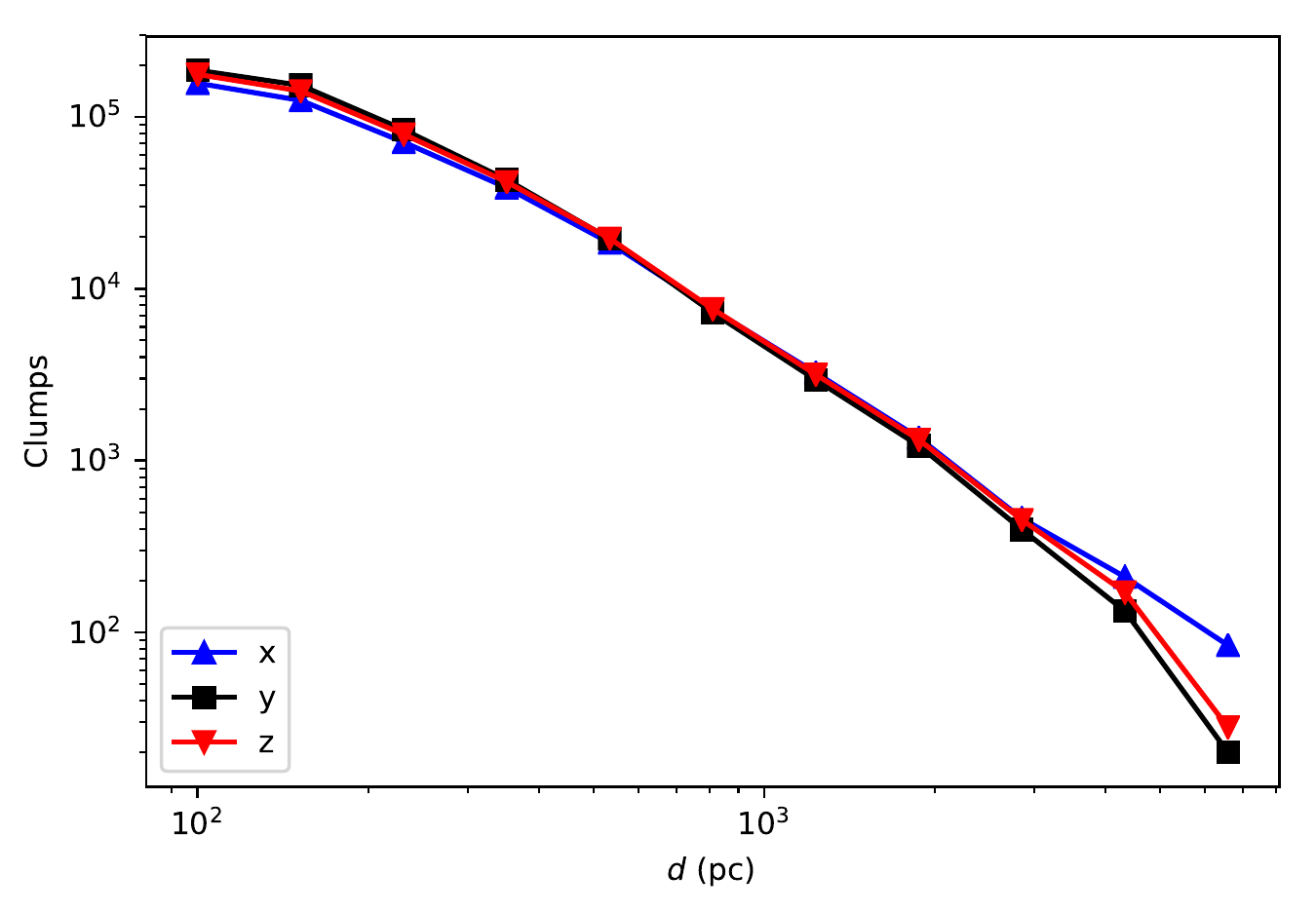}
\caption{
Number of extracted clumps as a function of cloud distance. The curves
correspond to the three different view directions and show the total
number of clumps in the 18 snapshots.
}
\label{fig:count_clumps}
\end{figure}

The total number of extracted clumps over all snapshots, view
directions, and distances is of the order of 1.5 million. In the
following we concentrate on sources with good photometry (S/N above one) in
all the four bands, which corresponds to the criterion FLUX\_QUALITY=1
in the PGCC\citep{PGCC}. Figure~\ref{fig:count_clumps} shows the number of
clumps as a function of distance. At the distance of $d=$100\,pc,
there are about 200 000 clumps per view direction, which corresponds
to some 11000 clumps per a single map (one snapshot and view
direction).

\begin{figure}
\includegraphics[width=8.8cm]{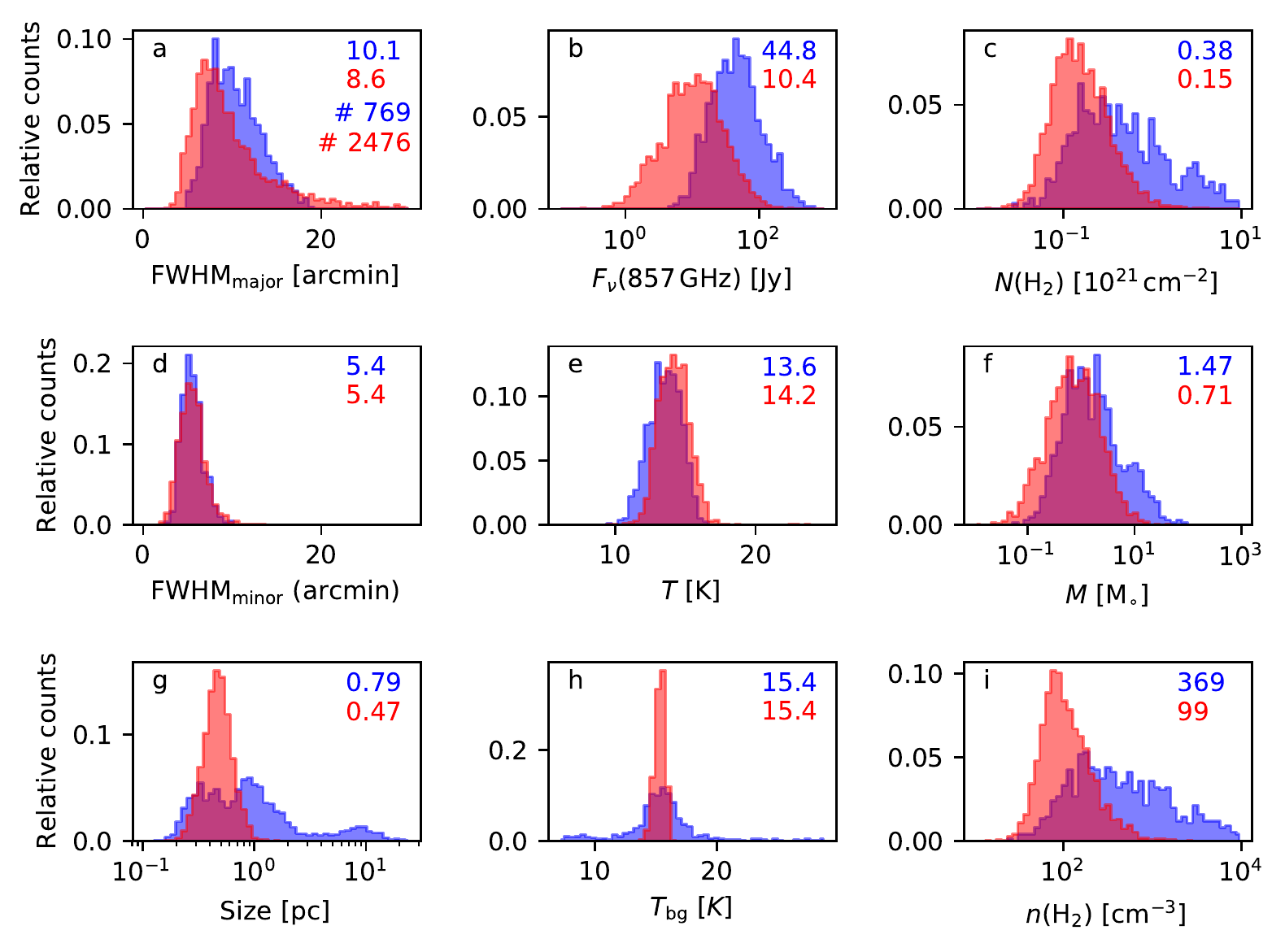}
\caption{
Normalised parameter distributions for clumps in snapshot 377,
direction $x$ and distance 231\,pc (red histograms). The blue
histograms show the corresponding distributions for PGCC sources with
within 50\% the same distance. The mean values of the parameters are
shown for the PGCC catalogue (upper numbers in blue) and for the
simulated catalogue (lower numbers in red). The first frame also shows
the number of clumps included in the plots.
}
\label{fig:377_x_231}
\end{figure}

Figure~\ref{fig:377_x_231} shows clump parameters for the snapshot
377, view direction $x$, and the distance of $d$=231\,pc. The
simulated clumps are compared to the PGCC sources that are to within 50\%
at the same distance. Both samples are limited to sources with
reliable photometry (in PGCC FLUX\_QUALITY=1). The FWHM sizes
correspond to the Gaussian fits. The linear sizes in pc are based on
the angular sizes and the known or, in the case of PGCC, estimated
distances.

At the shown distance, the synthetic maps provide a few times more
detections than the PGCC catalogue limited to similar distances within
$\pm 50\%$. Many clump parameters are comparable between the PGCC and the
synthetic catalogues. The minor axis sizes have almost identical
distributions, as a direct consequence of the common threshold set by
the beam size. The major axis angular sizes tend to be smaller for the
synthetic clumps but this is less clear for the physical sizes since
the PGCC distribution corresponds to a wider distance interval. The
clump sizes are similar for all view directions and thus independent
of the mean magnetic-field direction.

The largest discrepancy is in the flux densities that are about a
factor of four times lower for synthetic clumps. This is reflected
also in the estimates of the column densities, masses, and volume
densities. On the other hand, the temperature distributions are very
similar. The clump temperatures are based on the cold residual, not
the total dust emission, and the separation may contribute to the
similarity of the results. The background temperatures are also
similar, although the synthetic observations show a narrow
distribution because all models were subjected to the same radiation
field.

\begin{figure}
\includegraphics[width=8.8cm]{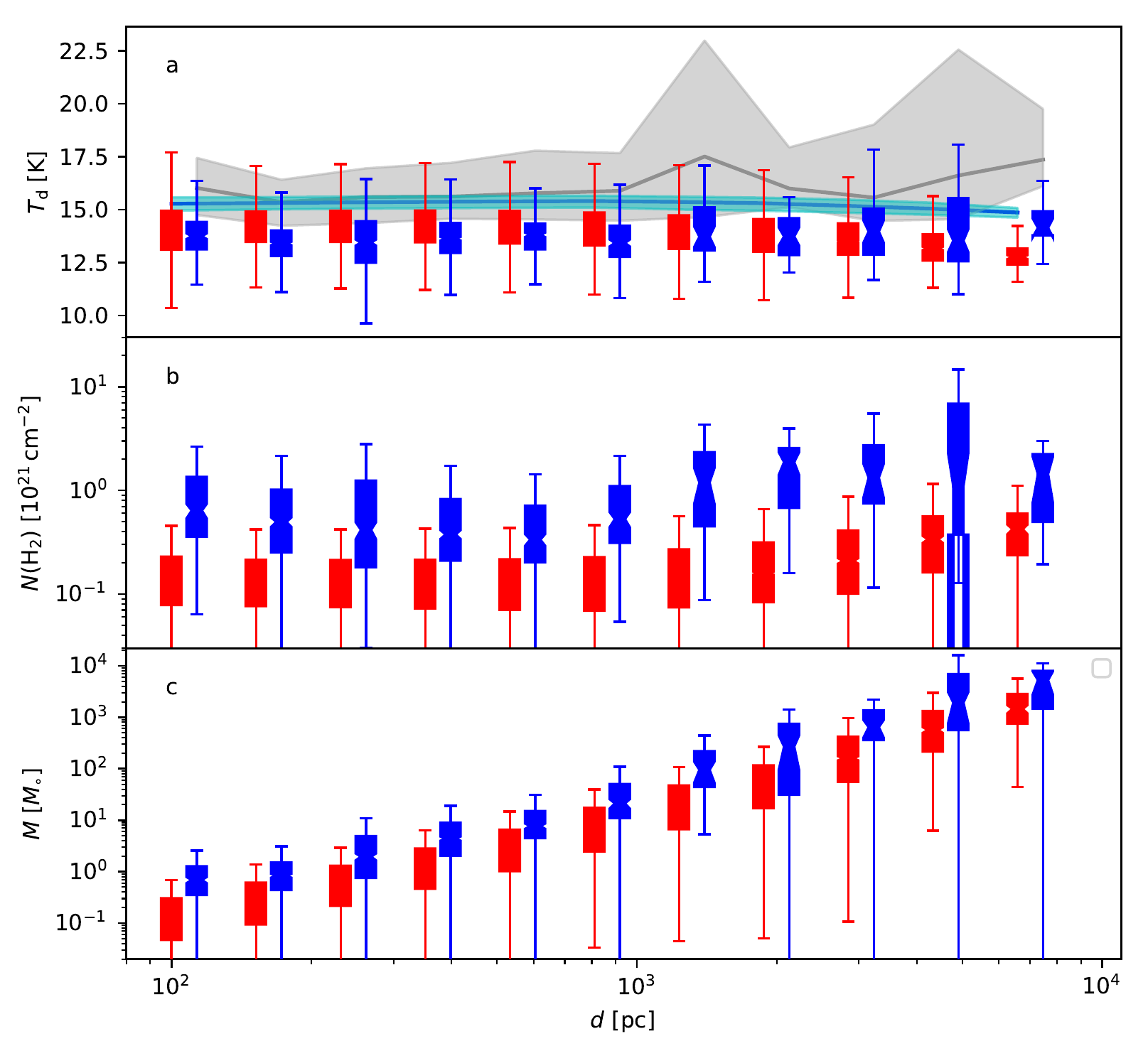}
\caption{
Comparison of clump temperature, column density, and mass as functions
of cloud distance. The red boxplots correspond to the simulated sample
and the blue boxplots to the PGCC. In Frame a, the grey shading and
the cyan shading indicate the interquartile intervals of the
background temperature in the PGCC and in simulations, respectively.
The median values are drawn with solid lines.
}
\label{fig:TNM_vs_distance}
\end{figure}

Figure~\ref{fig:TNM_vs_distance} shows the distributions of clump
temperatures, column densities, and masses as a function of distance.
This combines the statistics from all MHD snapshots and view
directions. The values from the PGCC catalogue with a similar distance
binning are shown for comparison.

Figure~\ref{fig:377_x_231} indicated that the model clump temperatures
are very similar to the PGCC values. Figure~\ref{fig:TNM_vs_distance}a
shows that this holds for all distances $d$ and that the average
temperatures do not depend on the distance. This is somewhat
unexpected given the two orders of magnitude difference in the probed
linear scales. However, the PGCC shows a similar behaviour and only a
marginal increase in the clump and background temperatures at the
largest distances (typically sources in the inner Galaxy). In the
simulations, this result could be expected because the radiation field
and thus the physical dust temperatures do not vary with distance.

Fig.~\ref{fig:TNM_vs_distance}b again shows the difference in the
column densities. The synthetic clumps have lower values but, as in
the case of the PGCC, there is a marginal increase at the largest
distances. The mass estimates are very strongly connected to the
resolution of the observations and scale as distance squared for both
the PGCC and synthetic observations.

\begin{figure}
\includegraphics[width=8.8cm]{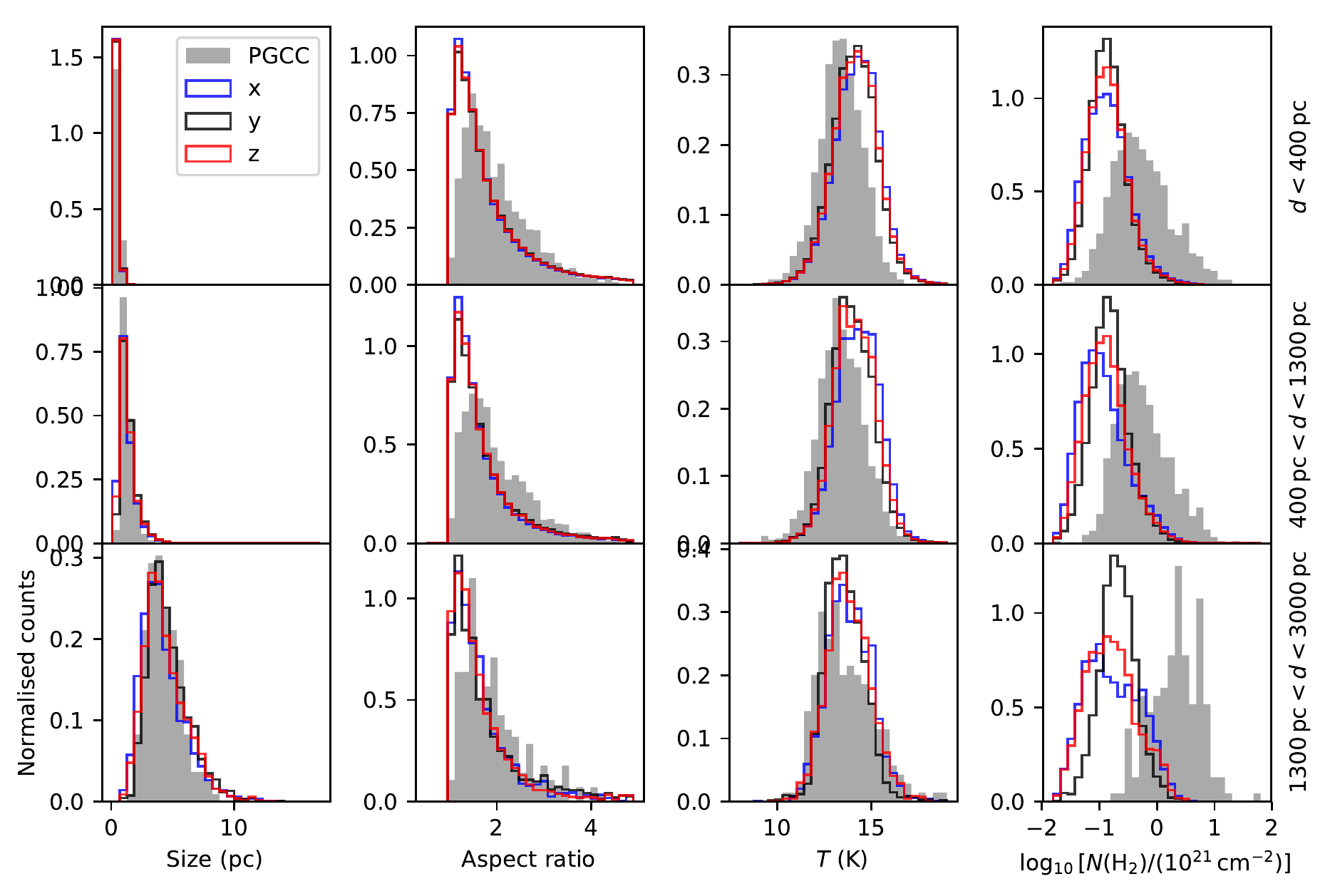}
\caption{
Comparison of synthetic clump parameters for three distance intervals
and view directions. The rows correspond to $d<400\,$pc, $400\,{\rm
pc}<d<1300\,{\rm pc}$, and $1300\,{\rm pc}<d<3000\,{\rm pc}$. The
columns show the distributions of physical clump size, aspect ratio,
colour temperature, and column density. The data for the different
view directions are shown in different colours. The corresponding data
from the PGCC are plotted as grey histograms. 
}
\label{fig:vs_direction}
\end{figure}

\begin{figure}
\includegraphics[width=8.8cm]{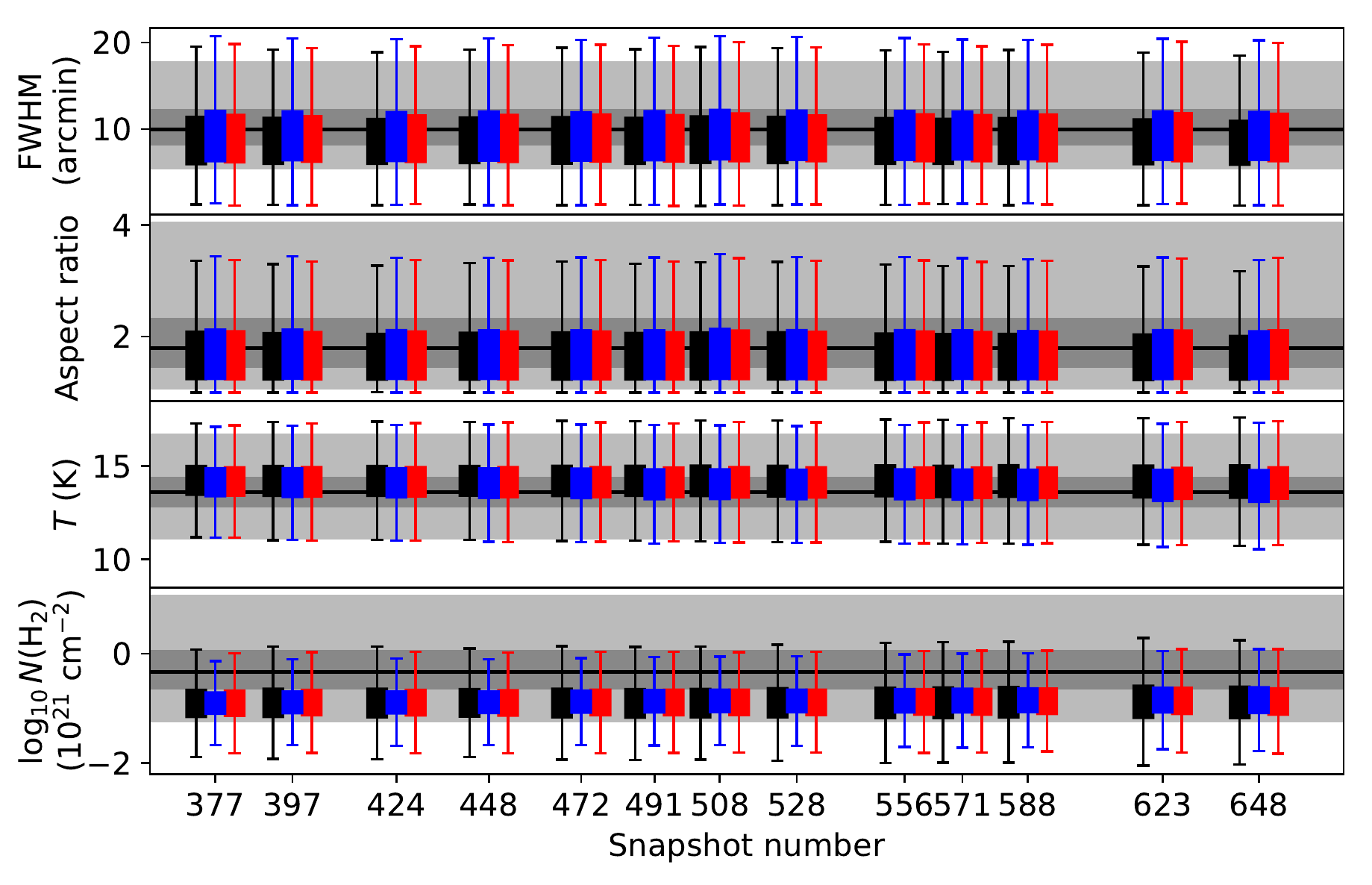}
\caption{
Distributions of clump FWHM, aspect ratio, temperature, and column
density as functions of MHD snapshot for clumps $d<400$\,pc. The
boxplots correspond to data for different snapshots and directions
(black, blue, and red for the $x$, $y$, and $z$ directions,
respectively). In each frame, the shaded regions correspond to the
distribution of the PGCC catalogue values, light grey for the 1\%-99\%
interval and dark grain for the 25\%-75\% interval of the cumulative
probability distribution.
}
\label{fig:plot_vs_snapshot}
\end{figure}

In the MHD simulations the turbulence is not isotropic because the
magnetic field has a non-zero mean component along the $y$ axis.
Figure~\ref{fig:vs_direction} shows observations for the three view
direction, but does not indicate any significant dependence between
the view direction and any of the clump properties. Similarly,
Fig.~\ref{fig:plot_vs_snapshot} shows that the clump properties do not
show time dependence, apart from a very minor increase of the average
column density.

\begin{figure}
\includegraphics[width=8.8cm]{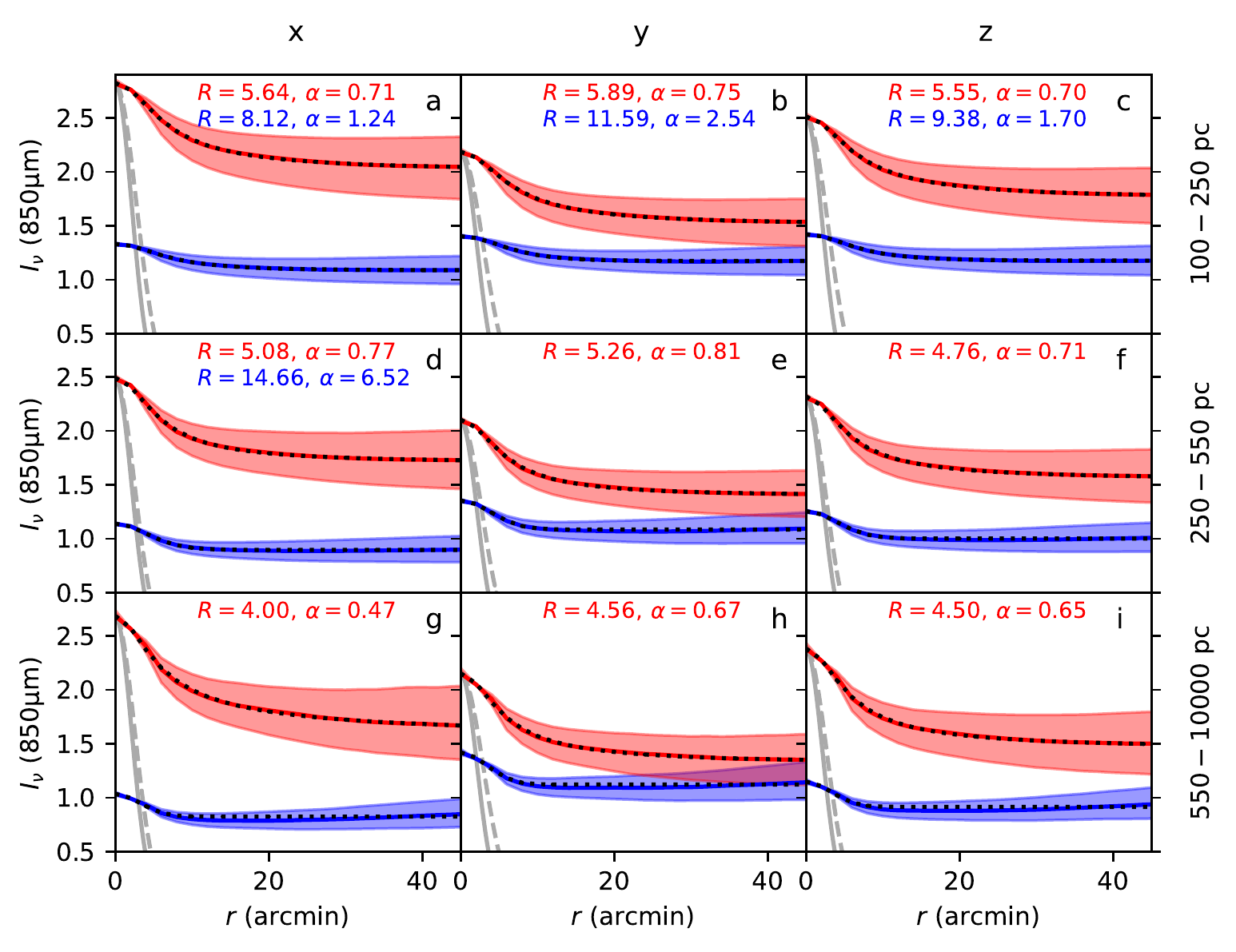}
\caption{
Overview of 850\,$\mu$m (353\,GHz) intensity profiles of 
extracted clumps, divided according to view direction (columns)
and distance (rows). Clumps with the central intensity above (red
curves) and below (blue curves) the median intensity are plotted
separately. The values are total intensities without background
subtraction, normalised to the sample median value at the centre. The
shaded regions correspond to the inter-quartile range. The median
profiles are shown as dotted lines with the parameters $R$ and
$\alpha$ (see text) quoted in the frames. The Gaussians corresponding
to the FWHM=5.0$\arcmin$ beam and the median major axis FWHM of the
extracted clumps (cold component) are plotted with solid and dashed
grey lines, respectively.
}
\label{fig:profiles_I}
\end{figure}

Figure~\ref{fig:profiles_I} shows median intensity profiles around the
extracted clumps. Clumps are divided into nine samples according to
the distance and view direction. There are further two sub-samples
according to the surface brightness at the clump centre. The FWHM
clump sizes tend to be only slightly larger than the beam. However,
those parameters correspond to the cold emission component while the
total intensity, as shown by Fig.~\ref{fig:profiles_I} is more
extended. All clumps detected in the cold residual emission are not
necessarily local maxima of total surface brightness, and the profiles
in Fig.~\ref{fig:profiles_I} thus characterise more the general mass
distribution in the clump environment than the intensity profile of
the clumps themselves. Figure~\ref{fig:profiles_I} includes only
clumps with reliable flux measurements (S/N above two), for which the
difference between the aperture and reference annulus is by definition
positive.

The median profiles of the total intensity are tentatively fitted with
functions
\begin{equation}
I(r) =  I_0  \left[ 1 + (r/R)^2\right]^{-\alpha} + I_{\rm bg}.
\label{eq:plummer}
\end{equation}
The parameter values $\alpha$ tend to be below $\alpha=1.0$ for the
brighter clumps, which means that the mean intensity falls off very
slowly, $\sim r^{-0.5}$. The fits to lower-intensity clumps are more
uncertain and for many low-intensity clumps the background level even
increase at large angular distances. Such profiles cannot be well
fitted with Eq.(\ref{eq:plummer}) and the fit parameters are not
shown.

\subsection{Polarisation fraction in basic models} \label{res:p}

In this section we examine the polarisation fraction $p$ of the clumps
and the correlation between $p$ and total intensity.
Section~\ref{sect:p1} concentrates on the analysis of the synthetic
observations and Sect.~\ref{sect:p2} looks at the correlations between
$p$ and the magnetic field structure.

\subsubsection{Synthetic polarisation observations} \label{sect:p1}

\begin{figure}
\includegraphics[width=8.8cm]{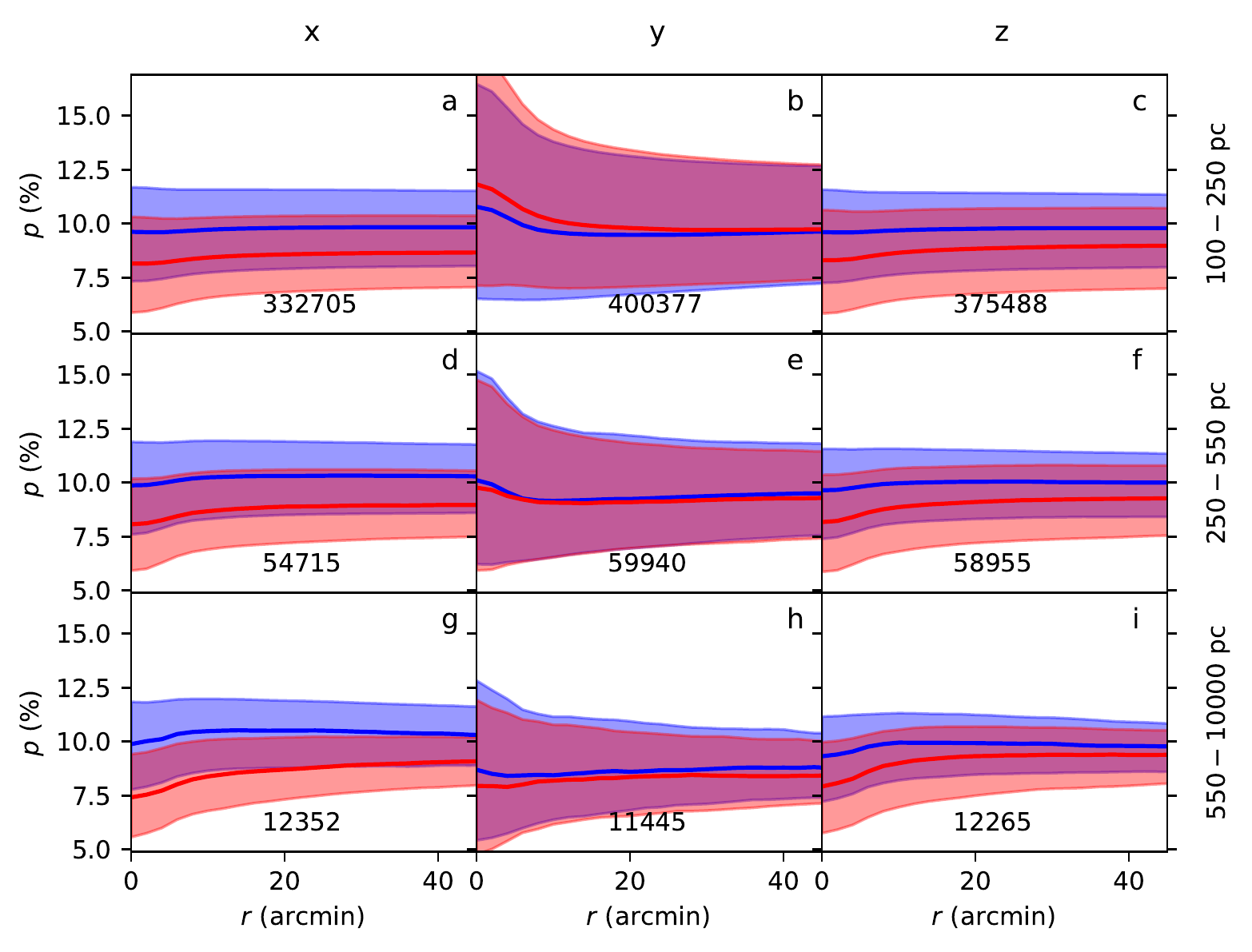}
\caption{
Radial profiles of median polarisation fraction $p$. The clump
sample is the same as in Fig.~\ref{fig:profiles_I}. The red and blue
colours correspond to clumps with the central surface brightness
respectively above or below the median value. The shaded regions show
the inter-quartile ranges. The total number of clumps is indicated in
each frame. The $y$-direction $p$ values have been multiplied by a
factor of four the plot.
}
\label{fig:profiles_p}
\end{figure}

Figure~\ref{fig:profiles_p} shows median polarisation fraction
profiles, similar to the intensity profiles of
Fig.~\ref{fig:profiles_I}. In the simulations the maximum theoretical
polarisation fraction was scaled to $p=20$\%. The values observed from
the models are typically below 10\%, especially in the dense regions
associated with clump detections. For the view directions $x$ and $z$,
the polarisation fraction tends to decrease towards the clump
positions. The drop is of the order of $\Delta p=1$\% and slightly
more pronounced at larger distances. The clumps are divided into two
sub-samples using the median value of the total intensity at the
centre of the clumps. Compared to the low-intensity clumps, the median
polarisation fraction of the high-intensity clumps is lower by $\Delta
p=$1-2\%. This applies to the $x$ and $z$ view directions and all
radial points, which means that this a property of the clump
environment rather than of the clump itself. Clumps thus reside in
dense regions where $p$ is already significantly below the average
over the entire model.

The behaviour is quite different for the $y$ direction where we
observe the cloud along the mean magnetic field direction. In this
case, the polarisation signal is much weaker (in the figure the values
are multiplied by a factor of four) and there is no clear difference
in the $p$ of the low-intensity and high-intensity clumps. For the
direction $y$, $p$ tends to increase rather than decrease towards the
clump centre. The behaviour is qualitatively consistent with the idea
of dense clumps locally perturbing the orientation of the large-scale
field, which would otherwise be parallel to the LOS.

\begin{figure}
\includegraphics[width=8.8cm]{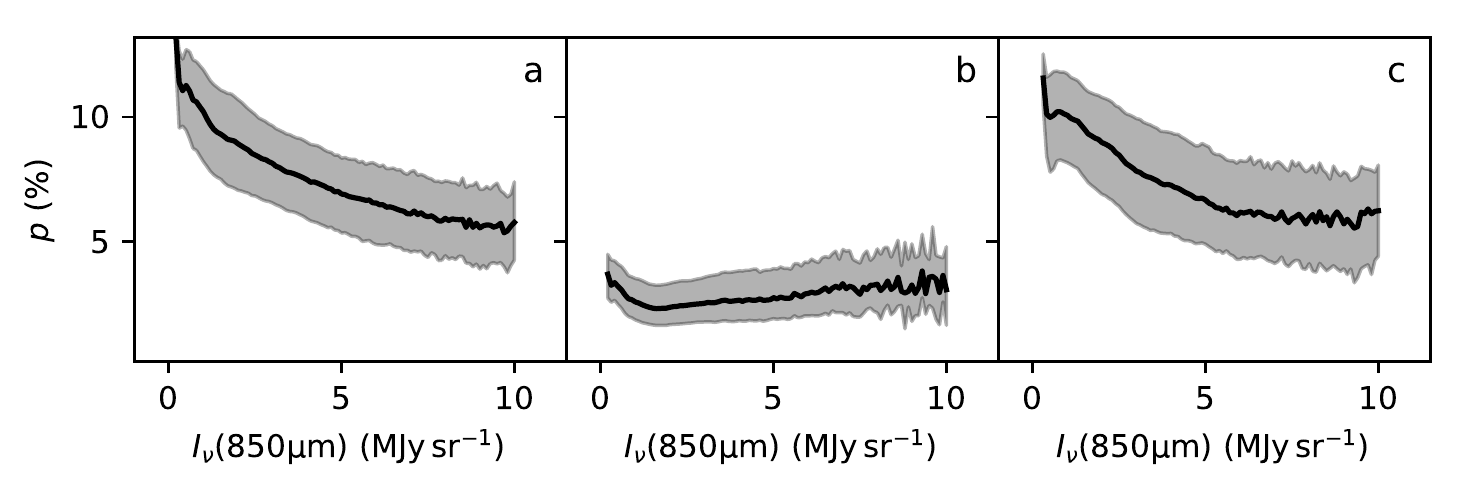}
\caption{
Polarisation fraction $p$ as a function of surface brightness for
clumps in Fig.~\ref{fig:profiles_I}a-c. The clump distances are $d \le
231$\,pc and the frames correspond to the view directions $x$, $y$,
and $z$, respectively. The shading indicates the inter-quartile range.
}
\label{fig:profiles_p_vs_I}
\end{figure}

The data from Fig.~\ref{fig:profiles_p}a-c are shown again in
Fig.~\ref{fig:profiles_p_vs_I}, now plotting $p$ against the surface
brightness $I_{\nu}$. The plot makes use of the azimuthal $p$ and
$I_{\nu}$ averages calculated up to a distance of 40$\arcmin$ of the
centre of each clump. For the directions $x$ and $z$, the decrease in
$p$ between the lowest and highest surface brightness areas is $\Delta
p \sim5$\% and thus almost a factor of two. For the direction $y$
(along the mean-field direction) the distribution is flatter and $p$
increases slightly towards both ends of the intensity axis. The $p$
peak at the clump centre, as seen in Fig.~\ref{fig:profiles_p}b, is
here stretched over intensities $2.5-10$\,MJy\,sr$^{-1}$.

Each frame of Fig.~\ref{fig:profiles_p} covers a heterogeneous set of
sources. Therefore, for the sources from Fig.~\ref{fig:profiles_p}a,c
we correlated $p$ with a number of other parameters, including the
background intensity, S/N of the detection, and the clump size, shape,
and temperature. For each clump, we also characterise the radial
change of $p$ with $\Delta p$, the difference between the mean values
at $0-4\arcmin$ and $10-16\arcmin$ radial distances. A negative value
of $\Delta p$ thus indicates a decrease of $p$ towards the clump
centre.

\begin{figure}
\includegraphics[width=8.8cm]{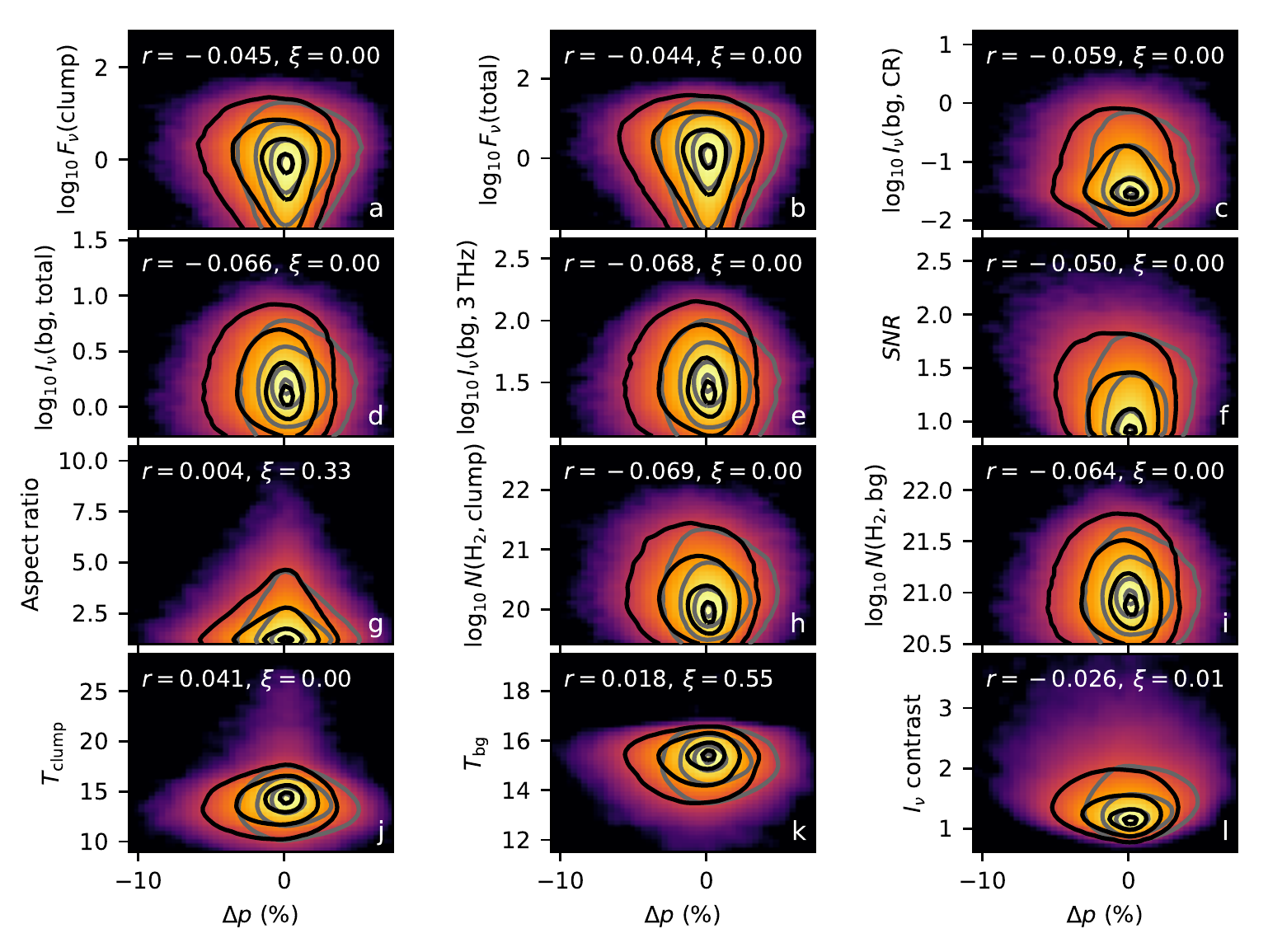}
\caption{
Correlations between various parameters and $\Delta p$, the drop in
polarisation towards centre of $d\le 231$\,pc clumps (see text).
The colour images show logarithmic point density for the clumps from
Fig.~\ref{fig:profiles_p}a-c. The frames quote the linear correlation
coefficients $r$. Probabilities $\xi$ for $r$ to be consistent with
zero are computed for a factor of 100 smaller random clump samples.
The black contours show the distribution for the combination of view
directions $x$ and $z$ and the grey contours for the direction $y$.
The contours are at 0.01, 0.1, 0.5, and 0.9 times the peak value. If
not otherwise specified, the intensities and flux densities are the
850\,$\mu$m values.
}
\label{fig:p_vs_parameters}
\end{figure}

Figure~\ref{fig:p_vs_parameters} shows kernel density estimates of
distributions when each parameter is plotted against $\Delta p$. The
overall correlations are weak with the absolute values of the linear
correlation coefficients $r$ below 0.2. The figure includes formal
probabilities $\xi$ for Pearson correlation coefficients to be
consistent with zero. Because of the set of correlated snapshots, the
probabilities could be biased and we therefore quote $\xi$ values that
are calculated for a factor of 100 smaller random clump samples. For
example, the comparison of $\Delta p$ and the corresponding intensity
contrast (the ratio of total 353\,GHz surface brightness at
$\theta<4\arcmin$ divided by the average value at distances
$\theta=10-16\arcmin$) gives $r=-0.026$; brighter clumps tend to have
lower values of $p$. The correlation coefficient is small but its
significance is still high.
Negative correlations with $\Delta p$ are observed also for the
detection S/N, source flux, source column density, and the background
in the cold residual map. The correlation is positive for the clump
and background temperatures. Clumps with very large aspect ratios
(possible filamentary morphologies) cluster around $\Delta p=0$.

\begin{figure}
\includegraphics[width=8.8cm]{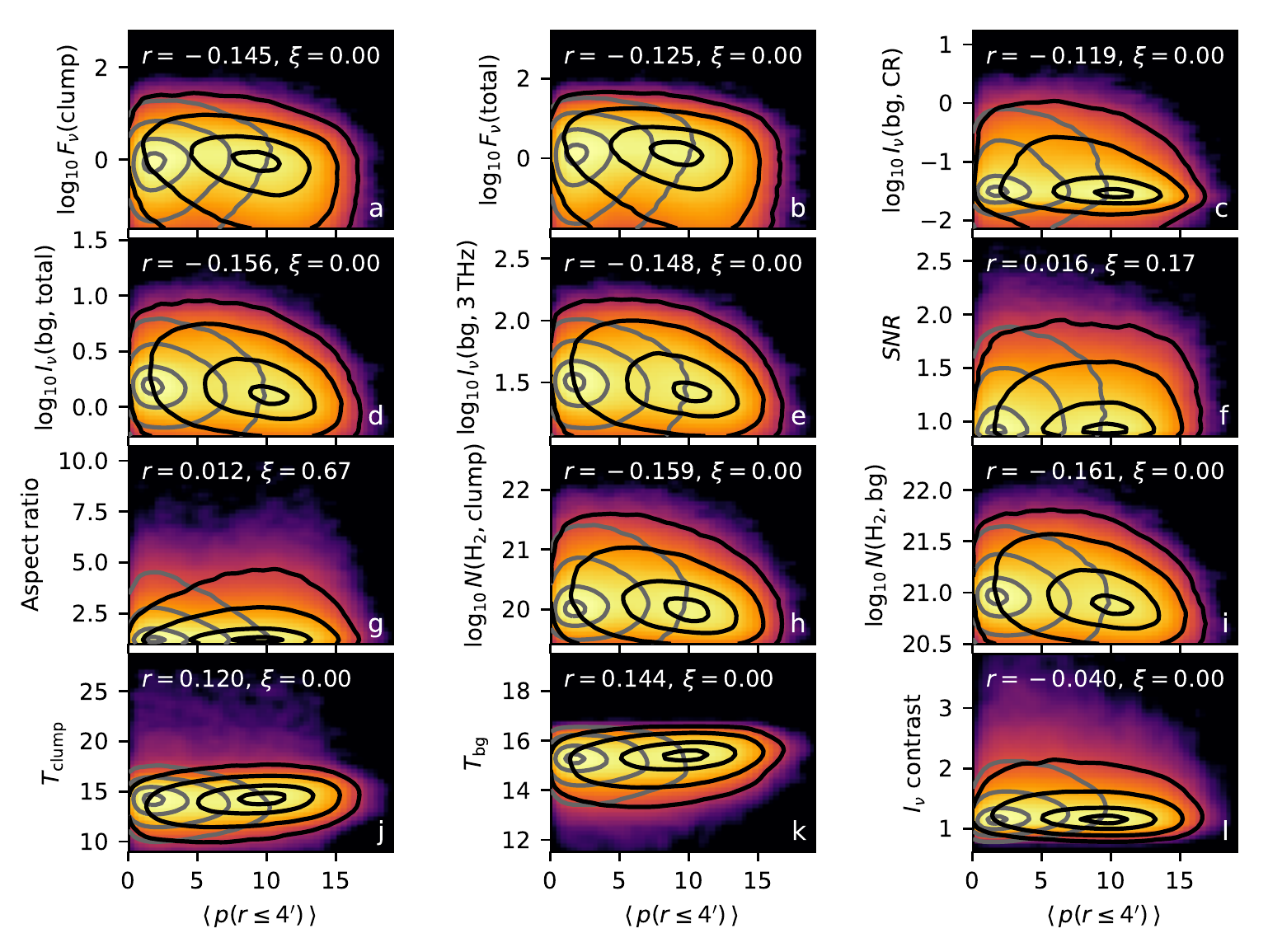}
\caption{
Same as Fig.~\ref{fig:p_vs_parameters} but plotting quantities against
$p$ at the clump centre.
}
\label{fig:p_vs_parameters_Pcentre}
\end{figure}

Figure~\ref{fig:p_vs_parameters_Pcentre} shows the corresponding
correlations when $\Delta p$ is replaced with the absolute value of
the polarisation fraction at the clump centre. This results in larger
correlation coefficients, which all retain the same sign as in
Fig.~\ref{fig:p_vs_parameters}. The only exception is the detection
S/N, for which the correlation is now positive but with a smaller
significance. Unlike in Fig.~\ref{fig:p_vs_parameters}, the
differences between the $y$ vs. $x$ and $z$ directions are clear.

\subsubsection{Comparison to magnetic field structure} \label{sect:p2}

\begin{figure*}
\sidecaption
\includegraphics[width=12cm]{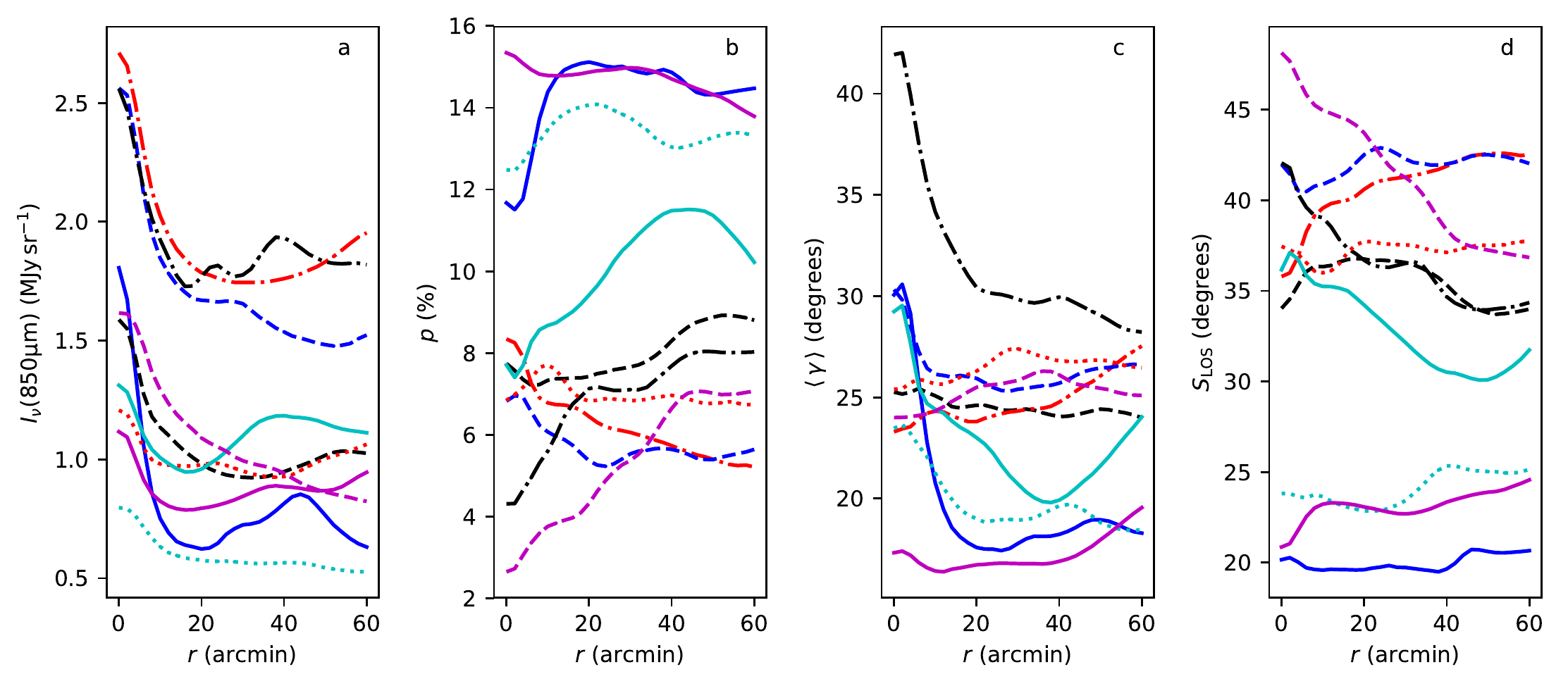}
\caption{
Examples of radial profiles of  $I_{\nu}(850\mu \rm m)$, $p$,
$\langle \cos^2 \gamma \rangle$, and $S_{\rm LOS}$. The ten sources
are selected randomly from the clumps in the snapshot 377, direction
$x$, distance 152\,pc, and ${\rm S/N}>15$.
}
\label{fig:sample_profiles}
\end{figure*}

The observed polarisation fraction variations are affected by the
magnetic field structure that can be characterised with $\langle
\,\gamma \, \rangle$, $S_{\rm LOS}$, and $S_{\rm POS}$
(Eqs.~(\ref{eq:gamma})-(\ref{eq:S_POS})).
Before discussing the statistics, we show in
Fig.~\ref{fig:sample_profiles} ten randomly selected clumps. There is
great variety in the type of $p$ profiles, polarisation sometimes
increasing rather than decreasing towards the clump centre. This small
sample already suggests that $p$ depends on both $\langle \, \gamma \,
\rangle$ and $S_{\rm LOS}$.

\begin{figure*}
\sidecaption
\includegraphics[width=12cm]{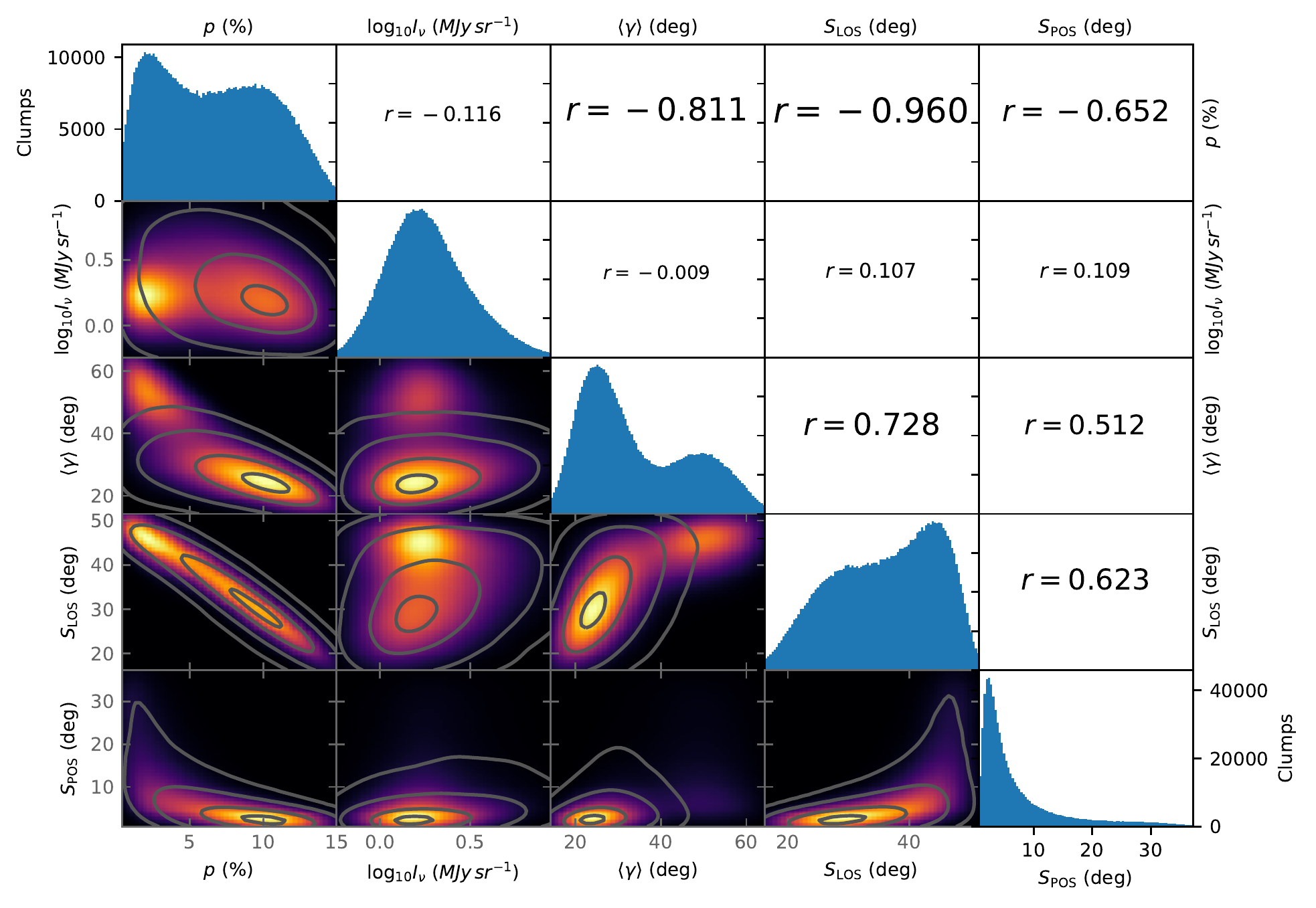}
\caption{
Correlations between parameters $p$, $I_{\nu}$, $\langle \gamma
\rangle$, $S_{\rm LOS}$, and $S_{\rm POS}$. The values are estimated
towards the centres of clumps at distances $d\le 231$\,pc. The
correlations are shown in the frames below the diagonal and the
histograms of the individual parameters on the diagonal. The upper
frames show the linear correlation coefficient $r$. The contours show
separately the distributions for the combination of $x$ and $z$ view
directions, with contours at 0.01, 0.1, 0.5, and 0.9 times the maximum
probability.
}
\label{fig:its}
\end{figure*}

Figure~\ref{fig:its} correlates the quantities for the full sample of
clumps at $d = 100-231$\,pc. Here $p$, $I_{\nu}$, $\langle \gamma
\rangle$ and $S_{\rm LOS}$ are estimated towards the centre of the
clumps while $S_{\rm POS}$ are averages within 10$\arcmin$ of the
clump centre.

The correlation coefficients show that $p$ is mostly dependent on
$S_{\rm LOS}$, $\langle \gamma \rangle$, and $S_{\rm POS}$, in that
order. All correlations are highly significant and especially the
dependence on $S_{\rm LOS}$ is very strong with $r=-0.96$. The formal
probabilities for $r$ to be consistent with zero are far below 1\%.
The parameters $\langle \gamma \rangle$, $S_{\rm LOS}$, and $S_{\rm
POS}$ show positive correlations also between themselves, with the
strongest one, $r=0.73$, between $\langle \gamma \rangle$ and $S_{\rm
LOS}$.

\begin{figure}
\sidecaption
\includegraphics[width=8.8cm]{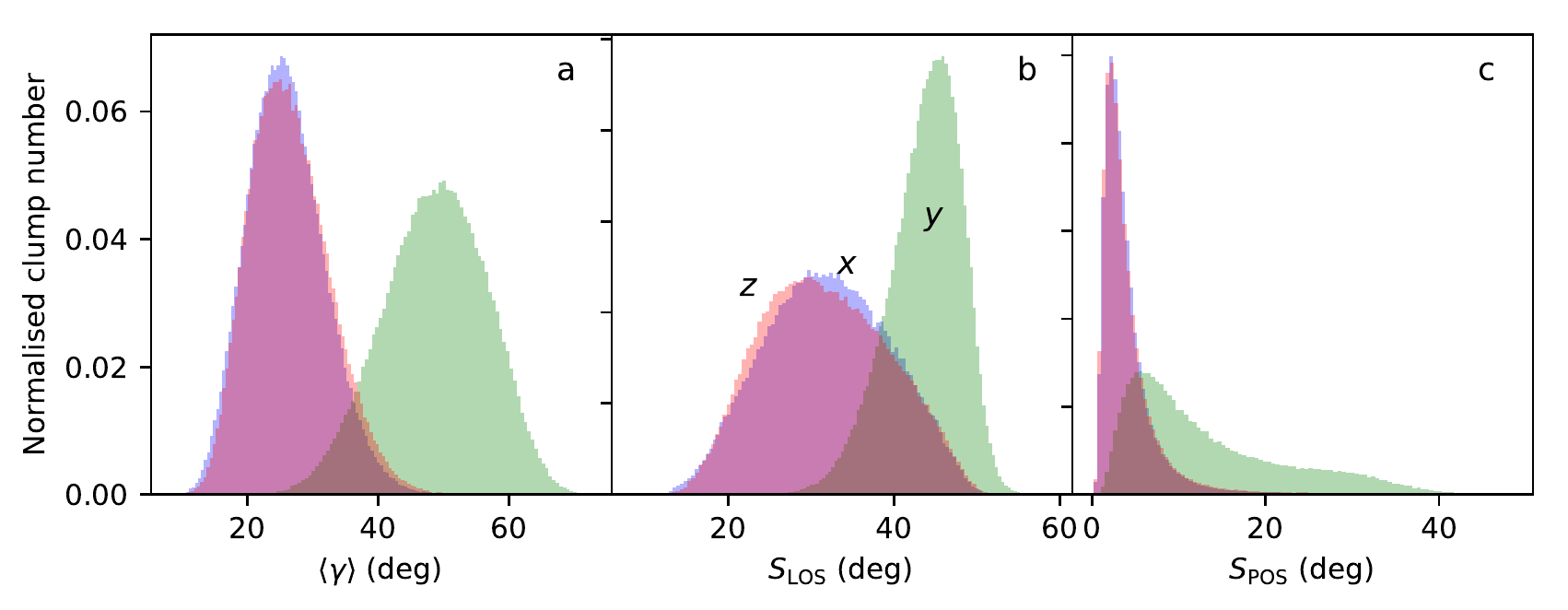}
\caption{
Histograms $\langle \gamma \rangle$, $S_{\rm LOS}$, and $S_{\rm POS}$
for clumps at $d=100-231$\,pc. The colours blue, green, and red
correspond to the view directions $x$, $y$, and $z$, respectively.
}
\label{fig:gamma_Slos}
\end{figure}

Some distributions of Fig.~\ref{fig:its} show two maxima that are
related to the different view directions. The direction $y$, being
parallel to the mean magnetic field orientation, is associated with
large $\langle \, \gamma \, \rangle$ and low $p$ values. The
importance of this effect is illustrated further in
Fig.~\ref{fig:gamma_Slos}, where we plot the histograms of $\langle
\gamma \rangle$, $S_{\rm LOS}$, and $S_{\rm POS}$ for the different
view directions.

Appendix~\ref{app:p} includes figures similar to Fig.~\ref{fig:its}
where the $y$ direction and the combination of the $x$ and $y$
directions are shown separately. While $S_{\rm LOS}$ always shows the
largest negative correlation with $p$, its importance is somewhat
reduced when the line of sight is parallel to the mean field.
Similarly, the correlations between $\langle \, \gamma \, \rangle$,
$S_{\rm LOS}$, and $S_{\rm POS}$ are lower for a given view direction.
This is expected because the sub-samples ($x$ and $z$ directions vs.
the $y$ direction) and internally more homogeneous. For the $y$
direction the parameters $\langle \gamma \rangle$ and $S_{\rm POS}$
become uncorrelated, although $\langle \gamma \rangle$ has significant
correlation with $S_{\rm LOS}$, which is even more correlated with
$S_{\rm POS}$.
Plots comparing polarisation data for different distances can also be
found in Appendix~\ref{app:p}.

\begin{table}
\caption[]{List of alternative models. For each view direction, the default
models are calculated for all 18 snapshots, models $L_2-L_4$ only for three
snapshot combinations, and the others for four selected snapshots.}
\begin{center}
\begin{tabular}{ll}
Symbol         &  Description  \\
\hline
\hline
$D$            &  default models                    \\
$L_2$, $L_3$, $L_4$  &  LOS longer by a factor of 2, 3, or 4 \\
$M_1$, $M_5$   &  higher dust opacity for $n(\rm H_2)>1000$\,cm$^{-3}$ \\
               &  or $n(\rm H_2)>5000$\,cm$^{-3}$, respectively  \\
$N$            &  observational noise higher by a factor of five \\
$H$            &  internal heating sources in the model volume \\
\hline
\end{tabular}
\end{center}
\label{table:models}
\end{table}

\subsection{Modified models}  \label{sect:mod}

We examined four modified versions of the synthetic observations.
These included (1) longer lines of sight, (2) modification of the dust
properties, (3) larger measurement noise, and (4) the inclusion of
point sources that increase radiation field variations within the model
volume (Table~\ref{table:models}).

\begin{figure}
\includegraphics[width=8.8cm]{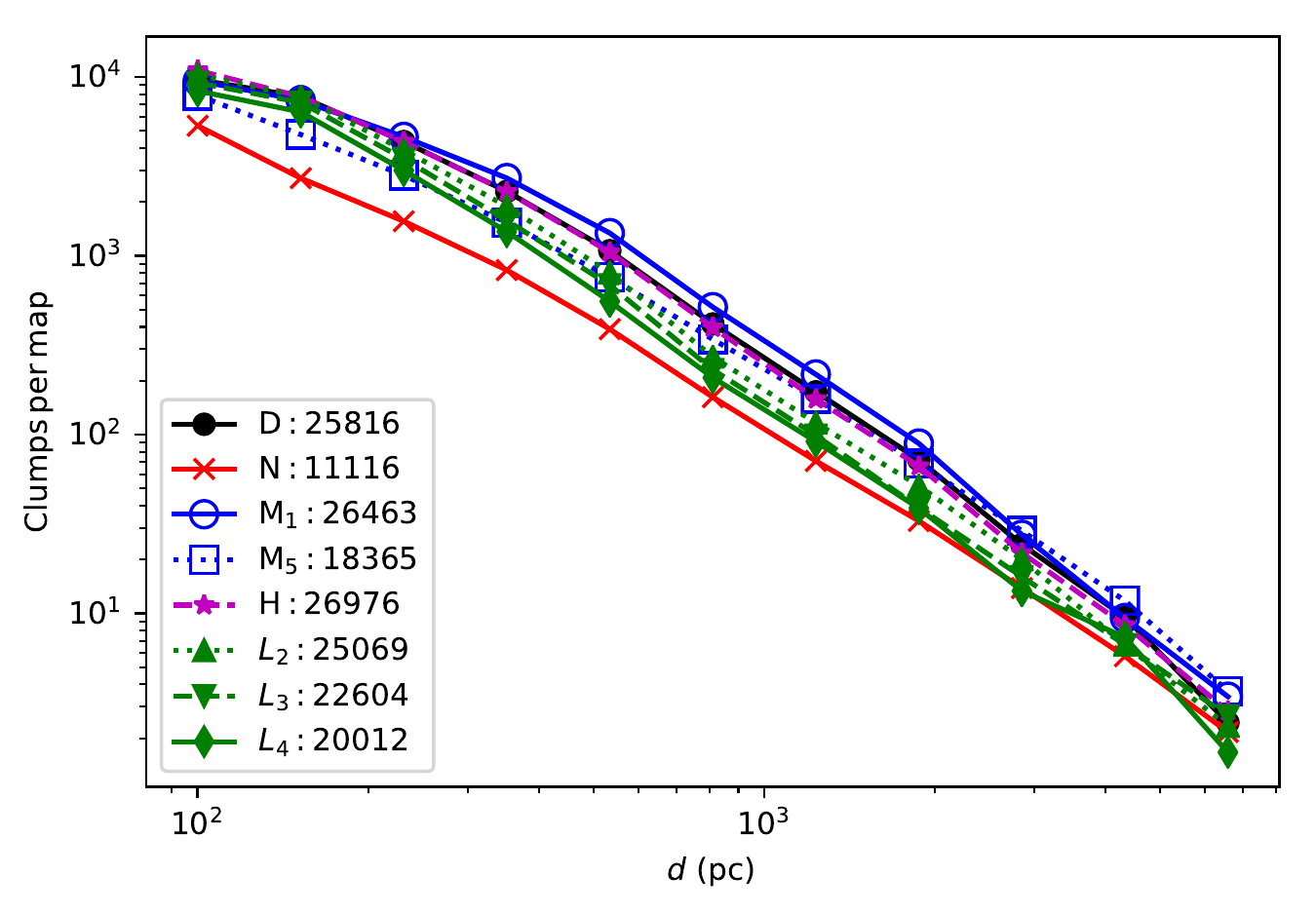}
\caption{
Number of clumps per map for alternative models. The legend
includes the average number of detected clumps per snapshot and
direction, as the sum over all distances.
}
\label{fig:compare_clump_counts}
\end{figure}

\begin{figure*}
\sidecaption
\includegraphics[width=12cm]{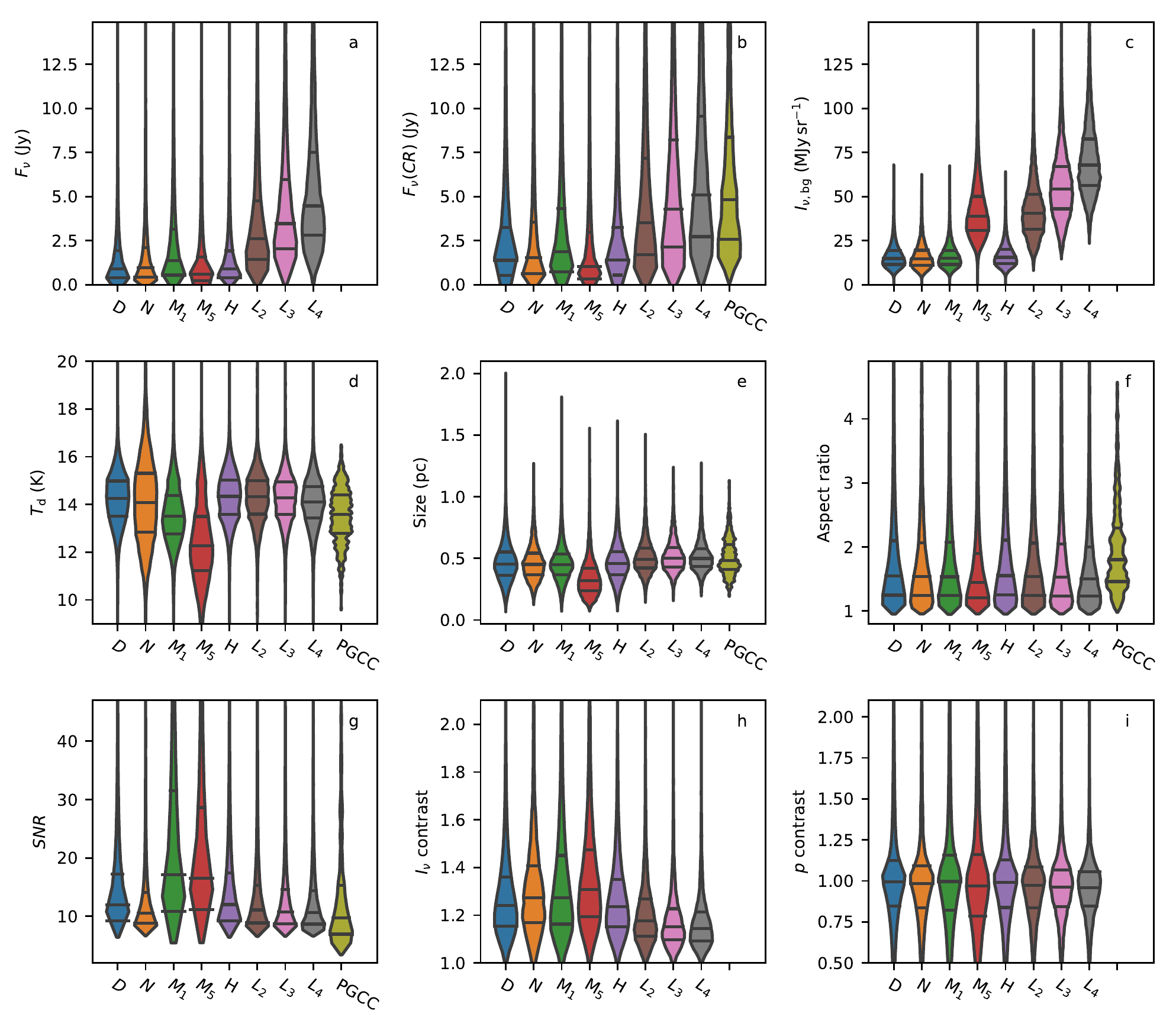}
\caption{
Comparison of parameters for alternative models. Each frame shows
parameter distributions for the models (see Table~\ref{table:models})
and for the PGCC. The distributions are plotted using kernel density
estimates (`violin plots') with the quartile points indicated with
horizontal lines. The frames a-i show: clump 850$\mu$m flux density,
850\,$\mu$m flux density in the cold residual map, 100\,$\mu$m
background intensity, clump temperature, physical size, and aspect
ratio, detection S/N, and the intensity and polarisation fraction
contrasts (calculated as ratios of the mean values at radiae
$R<4\arcmin$ and $R=10\arcmin-16\arcmin$). Model results are for the
distance of $d=231$\,pc. The PGCC distributions are plotted for selected
quantities, for clumps with FLUX\_QUALITY=1 and distance estimates
within a factor of 1.5 of the nominal distance.
}
\label{fig:plot_mod}
\end{figure*}

\subsubsection{Models with longer LOS} \label{sect:longer}

A longer effective LOS $L$ was obtained by adding together two or more
of the original surface brightness maps. Each sum contains two,
three, or four maps that thus correspond to $L=$2, 3, or 4 times
longer LOS. For each $L$, we made three sets of maps using different
snapshots and view directions. For example, $L=2$ used combinations
377$x$+424$y$, 443$y$+472$z$, and 491$z$+571$x$, where the numbers
refer to the snapshots and the letters to the view directions. For
$L=3$, each new map is the sum of three maps, all from different
snapshots and view directions. Finally, the $L=4$ maps are the sum of
four maps that thus contains the same direction ($x$, $y$, or $z$)
twice. Although these originate in different snapshots, the latter was
transposed before the summation to avoid undue correlations between
the component maps. For each $L$, in total exactly one third of the
input maps is for the $y$ direction.

\begin{figure}
\includegraphics[width=8.8cm]{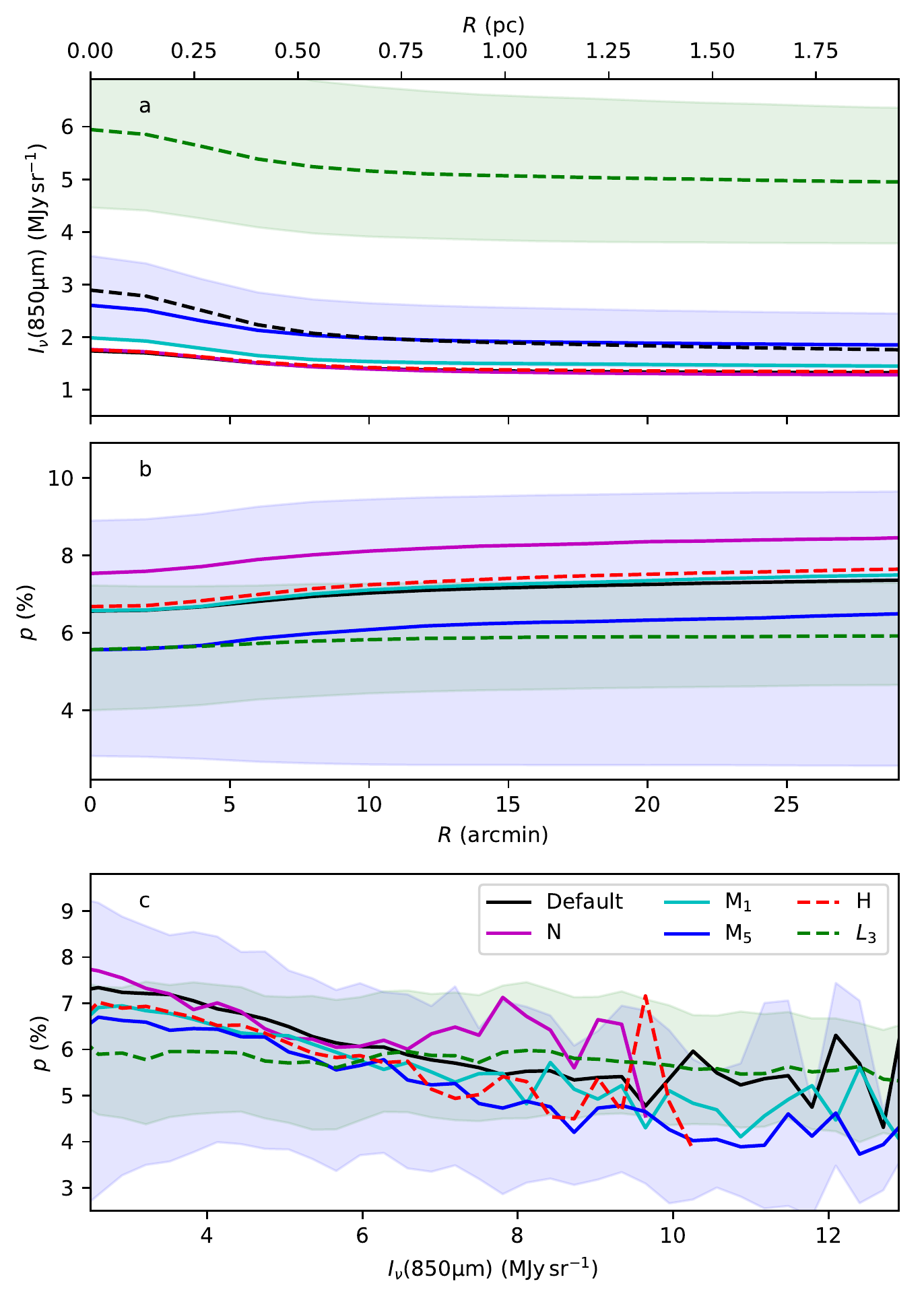}
\caption{
Radial profiles and correlation between polarisation fraction and
surface brightness for alternative models with $d=231$\,pc, as
indicated in the last frame. The shaded regions correspond to the
inter-quartile intervals for the models $L_3$ (green) and $M_5$
(blue). Frames a-b include additional curves for a ${\rm S/N}>10$, $T_{\rm
d}<13$\,K sub-sample of the default case (dashed black lines). In
frame b, these are on top of each other. In frame c, the curves for
the default and $H$ cases also overlap.
}
\label{fig:plot_mod_radial}
\end{figure}

Longer LOS reduces the number of clump detections
(Fig.~\ref{fig:compare_clump_counts}) 
but the fluxes of the extracted sources are higher and increase with
$L$ (Fig.~\ref{fig:plot_mod}). 
For the polarisation fraction profiles, the net effect of longer LOS
is a $\Delta p \sim1$\% reduction in the polarisation fraction
(Fig.~\ref{fig:plot_mod_radial}),
which tends to increase with increasing $L$ (not shown).  The relation
of $p$ vs. surface brightness also becomes flatter but this is mainly
the effect of the increased intensity values.

\begin{figure}
\sidecaption
\includegraphics[width=8.8cm]{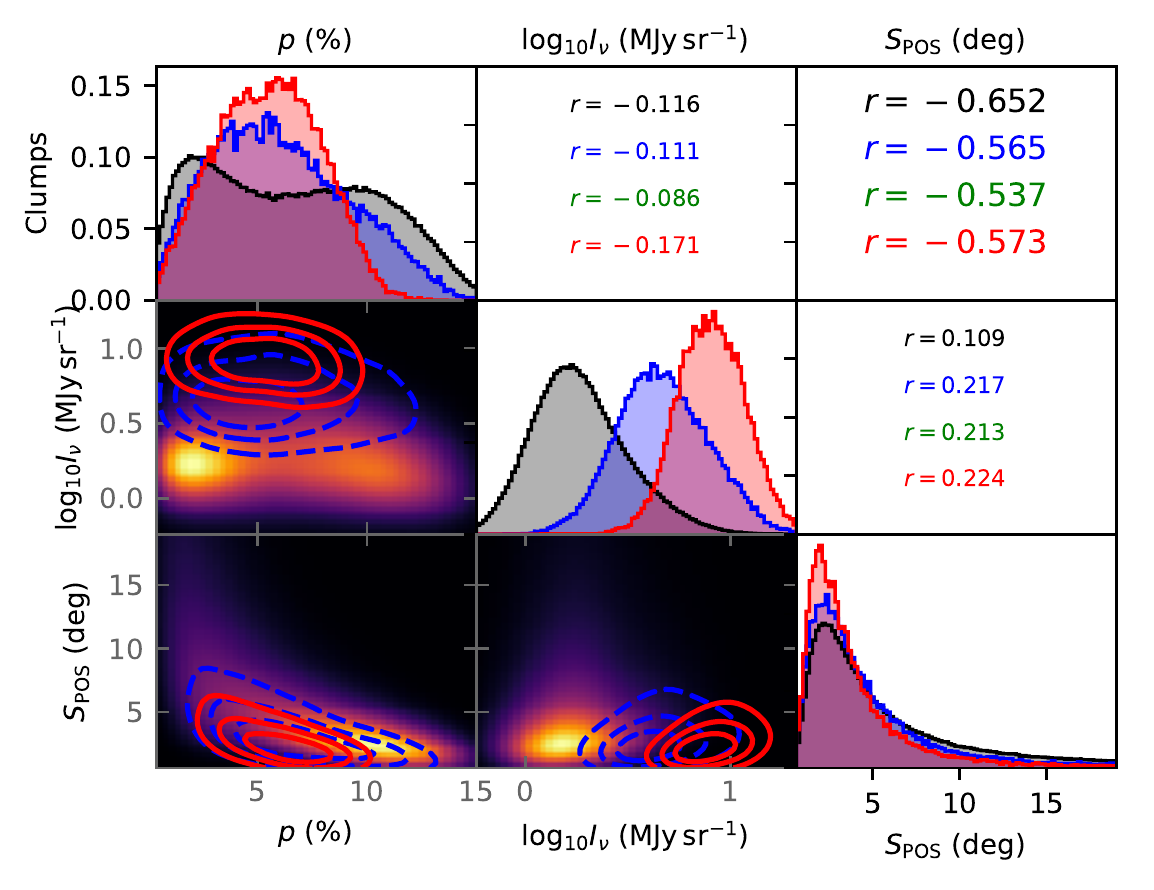}
\caption{
Comparison of $p$, clump 850\,$\mu$m intensity $I_{\nu}$, and $S_{\rm
POS}$ for default model and longer LOS cases. The background
images and the black histograms correspond to the default model and
the blue and red histograms, respectively, to the $L=$2 and $L=$4
models. From top to bottom, the correlation coefficients are listed
for the default case and the $L=$2, 3, and 4 cases. The plot includes
all clumps at $d\le 231$\,pc.
}
\label{fig:correlate_p_vs_ITS_LOS}
\end{figure}

Figure~\ref{fig:correlate_p_vs_ITS_LOS} compares $p$ to surface
brightness and polarisation angle dispersion function $S_{\rm POS}$.
The absolute surface brightness is obviously correlated with $L$.
Because of the mixing of different view directions $x$, $y$, and $z$
into single maps, the $p$ distribution shows here only a single peak
that is between the cases of LOS being either parallel or
perpendicular to the mean field. More interestingly, the
differences in $S_{\rm POS}$ distributions remain small and longer LOS
even shows a slight preferences for lower $S_{\rm POS}$.


\subsubsection{Dust properties} \label{sect:dust_properties}

Alternative models were created also by changing the dust properties.
Only snapshots 377, 406, 444, and 528 were used in these tests.

We consider a second dust component where the ratio of sub-millimetre
and NIR dust opacities is $\tau(250\mu {\rm })/\tau(J)=1.6\times
10^{-3}$, as derived from the statistical study of the PGCC fields
observed with Herschel \citep{GCC-V}. This is a factor of 3.2 increase
of dust opacity compared to the default dust model and this relative
rise was applied to $\lambda>30\mu{\rm m}$. The abundance of this
second dust component is
\begin{equation}
\chi = \frac{1}{2} 
+ \frac{1}{2} \tanh ( 2 \ln ( n({\rm H})/n_0) ).
\end{equation}
The sum of the default and the modified dust components was kept
constant. Therefore, at low densities the dust properties are as in
the default model and the increased-opacity dust is limited to regions
with high densities. We tested the threshold values of $n_0=1000\,{\rm
cm}^{-3}$ and $n_0=5000\,{\rm cm}^{-3}$ and refer to these models as
$M_1$ and $M_5$, respectively.

Figure~\ref{fig:plot_mod} shows that in the modified dust case the
clumps tend to have higher S/N and lower temperatures. The temperature
drop is particularly clear for the higher density threshold (models
$M_5$) where the median value is close to $T_{\rm d}=12$\,K. The
background intensities of $M_5$ clumps are also much higher while the
fluxes of the clumps themselves are even slightly lower.  The opacity
increase is thus not solving the difference to the higher PGCC source
fluxes. The density threshold should affect the size of the clumps, at
least if it corresponds to a volume with the size close to the
original size of the extracted sources. The higher threshold of
$n({\rm H})=5000\,{\rm cm}^{-3}$ indeed leads to smaller clumps but
the lower threshold of $n({\rm H})=1000\,{\rm cm}^{-3}$ has no
appreciable effect. The possible explanations are discussed in
Sect.~\ref{sect:alt}.

Figure\ref{fig:plot_mod_radial}a shows that the $M_5$ clumps tend to
have a high surface brightness. This means that the low flux densities
result from the smaller size of the clumps. Model $M_5$ clumps also
tend to show lower polarisation but with similar shape of the radial
profiles as the default models. Compared to $M_5$, the changes in
$M_1$ are in the same direction but much less pronounced. This is
interesting given that in $M_1$ the dust property changes extend over
larger areas. More than modifying the properties of individual clumps,
the dust changes influence which clumps are detected.

The effect of dust properties on the observed polarisation can be seen
also in Fig.~\ref{fig:plot_mod_pol}. The distribution of $p$ is more
skewed towards small $p$ values and the distribution of $S_{\rm LOS}$
is correspondingly skewed towards larger angle dispersion values.

\subsubsection{Increased noise}

We tested the effects of noise using the same set of snapshots as in
Sect.~\ref{sect:dust_properties}. The noise of the surface brightness
maps (see Sect.~\ref{met:RT}) was increased by a factor of five.
Higher noise reduces the number of clump detections by almost the same
factor (Fig.~\ref{fig:compare_clump_counts}) and increases especially
the dispersion of the clump temperature estimates. The noise also
reduces the flux estimates of the clumps (the cold component)
(Fig.~\ref{fig:plot_mod}). Figure~\ref{fig:plot_mod_radial} shows that
the noise has increased the clump polarisation on average by $\Delta
p\sim 1$\%. Because the polarisation fraction was calculated from data
without any added noise, this is not caused by bias in the $p$ values
but is a result of changes in the clump detection.

\subsubsection{Radiation field variations}

\begin{figure}
\includegraphics[width=8.8cm]{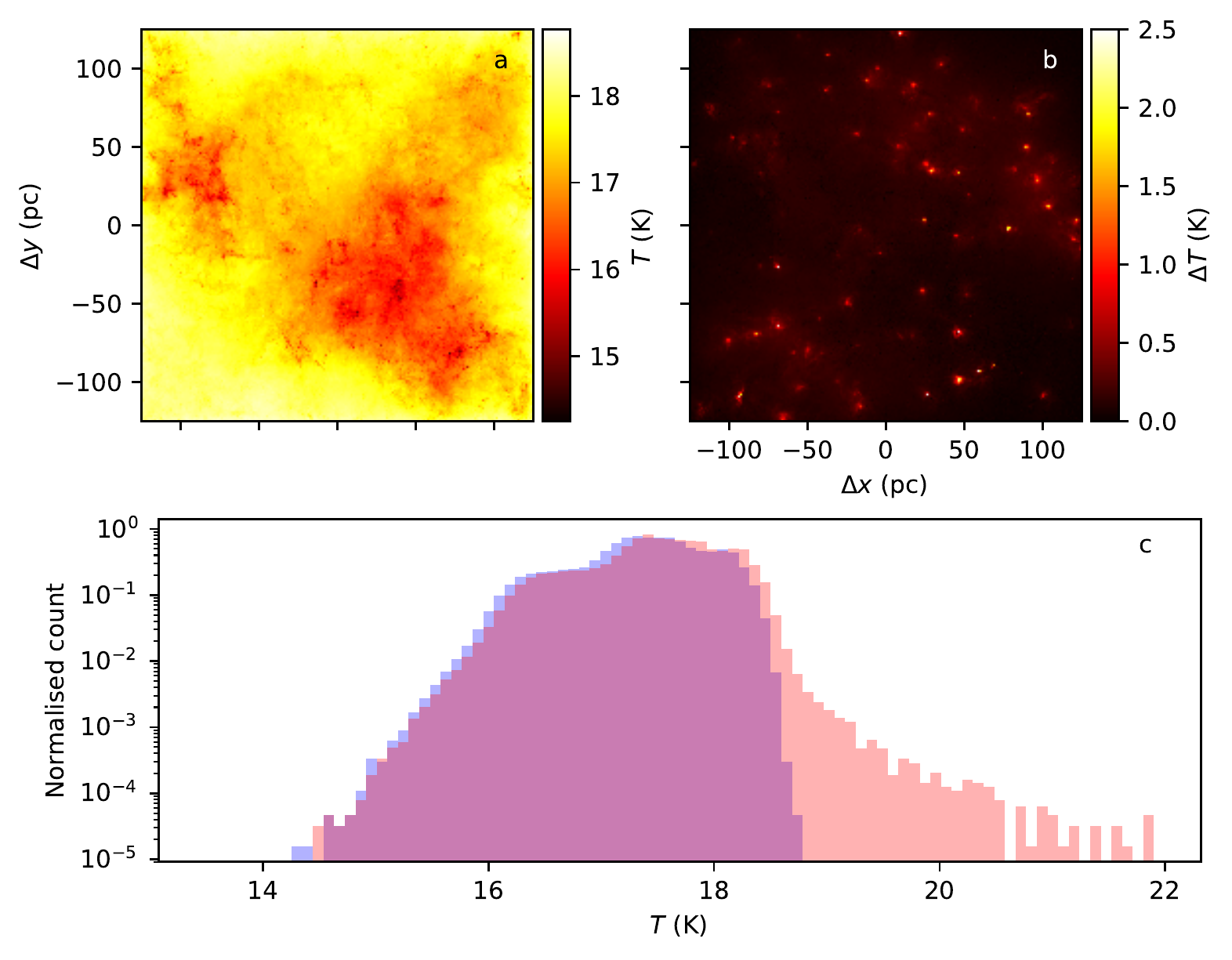}
\caption{
Effect of internal sources on dust colour temperature. Frame a
shows the temperature map for the default model $D$ (snapshot 406,
direction $x$) and frame b the change in the temperatures resulting
in model $H$ from 100 radiation sources inside the model volume. The
temperature histograms are shown in frame c (blue for model $D$ and
red for model $H$).
}
\label{fig:plot_star_Tmap}
\end{figure}

In the final set of alternative models, we added heating sources
inside the model volume. This was done not to simulate protostellar
cores, but to induce variations in the radiation field intensity and
to investigate the effects that the resulting variations of the
background temperature have on clump extraction. After all, according
to Fig.~\ref{fig:377_x_231}h, one of the main differences compared to
the PGCC catalogue is the uniformity of the colour temperature in the
clump background. One hundred point sources were added to each of the
four selected snapshots (377, 406, 444, and 528). The positions of the
sources were selected randomly with uniform probability over the model
volume. The sources were described as $T=$20000\,K black bodies, with
luminosities sampled from a normal distribution $N$(1000\,$L_{\sun}$,
300\,$L_{\sun}$). The luminosities are high enough to increase even
the long-wavelength surface brightness by tens of per cent over
projected distances of the order of 10\,pc. Figure
~\ref{fig:plot_star_Tmap} shows the changes in dust colour
temperature. 

\begin{figure*}
\sidecaption
\includegraphics[width=12cm]{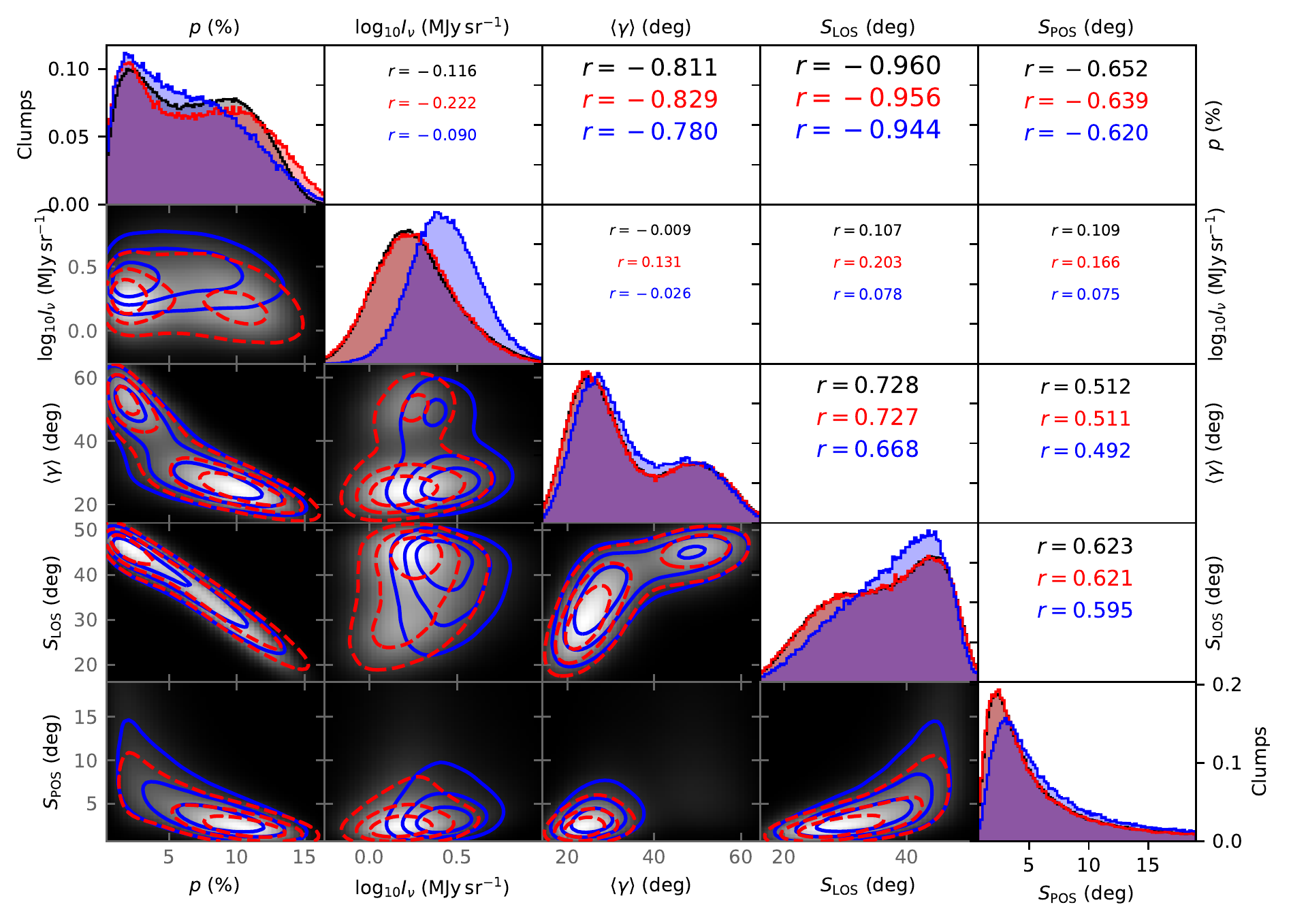}
\caption{
Comparison of polarisation-related parameters for alternative models.
The lower frames show kernel-density-estimated parameter correlations
and the diagonal frames the histograms of individual parameters. The
correlation coefficients are listed in the remaining frames. The
background images and histograms drawn with black lines correspond to
the default model $D$ with all clumps $d<300$\,pc. The data for the
alternative models $H$ (internal heating sources) and $M_5$ (modified
dust properties) are drawn in red and blue, respectively.
}
\label{fig:plot_mod_pol}
\end{figure*}

Although the effect of the heating sources is clear in the appearance
of the temperature maps, it still affects significantly only a small
fraction of the map areas. The effects on the parameters of the
extracted clumps remain small (Fig.~\ref{fig:plot_mod}). The same
applies to polarisation quantities where the changes with respect to
the default model $D$ are smaller than for the alternative dust models
$M_5$, with only a marginal widening of the $p$ distribution
(Fig.~\ref{fig:plot_mod_pol}). The $p$ distribution has a stronger
peak at small values but the other parameters do not shown 
significant changes.

\section{Discussion} \label{sect:discussion}

We used radiative transfer modelling and cold-source
detection methods to study the properties of dense clumps in MHD
simulations of molecular clouds. The results were compared to the
PGCC catalogue \citep{PGCC}. Below we discuss the clump extraction,
the clump characteristics, and the relationships between the clump
polarisation and the model clouds.

\subsection{Clump detection}

The detection and source analysis methods followed the example of the
PGCC catalogue \citep{PGCC}. The detection method itself was
already characterised in \citet{Montier2010} and in
\citet{planck2011-7.7b}.
Because the sources are extracted from the cold residual maps, their
temperature should be significantly lower than the temperature of
their environment. The  main physical explanation for
lower-than-average temperatures is high column density that
reduces the local radiation field. This is particularly true in our
simulations that do not include YSOs that could heat the clumps from
inside.  Nearby hot sources could lead to false detections by
biasing the background temperature estimates. This necessitated the
implementation of further safeguards in the analysis of the real
observations \citep{PGCC} but does not impact our simulations (without
embedded sources).

At the closest distances the number of extracted clumps was of the
order of 10000 per map (Fig.~\ref{fig:count_clumps},
\ref{fig:compare_clump_counts}), similar to the number of sources in
the PGCC catalogue. Although the models cover only a (250\,pc)$^3$ 
volume, this is reasonable because they represent relatively dense ISM
($\langle n({\rm H}_2) \rangle=5$\,cm$^{-3}$) and the number of
resolution elements in synthetic maps is up to 50\% of that in the
Planck all-sky map. However, while in the PGCC only half of the
detections had reliable fluxes, in the synthetic observations the
corresponding fraction is over 90\%. This is partly due to the absence
of internal sources. Our synthetic observations also suffer less from
LOS confusion, especially when compared to the PGCC where most
detections are near the Galactic plane. 


\subsection{Clump parameters in the default models}

Most clump properties were similar to those in the PGCC catalogue. The
almost identical distributions of minor axis FWHM reflect the cut-off
set by the beam size. The major axis FWHM values are comparable
although, in the basic models, the synthetic clumps tend to be smaller
(Fig.~\ref{fig:377_x_231}, \ref{fig:vs_direction},
\ref{fig:plot_vs_snapshot}). The size distribution of the PGCC is wider
(Fig.~\ref{fig:377_x_231}) also because they include a range of
estimated distances and the estimates have $\sim$30\% uncertainty
\citep{PGCC, GCC-IV}.

For the default models, the most noticeable difference between
synthetic and PGCC clumps is in the flux densities. In
Fig.~\ref{fig:377_x_231}, the PGCC clumps are on average about four
times brighter and this difference is carried over to column density,
mass, and volume density. 

The clump temperatures are slightly higher than in the PGCC, but this
finding of course depends on the assumed radiation field and the dust
model. We simulated only the large-grain emission but the contribution
to the 100\,$\mu$m band from stochastically heated very small grains
(VSGs) should get mostly eliminated when the warm-emission component
is subtracted. If VSG emission caused limb brightening in the real PGCC
clumps, their cold residual emission would be estimated to be smaller
in their outer parts, thus leading to smaller clump sizes and flux
densities. This is contrary to our finding that the synthetic clumps
tended to be smaller.

\subsection{Alternative cloud models} \label{sect:alt}

The default dust model was consistent with diffuse regions
(Sect.~\ref{met:RT}), with $\tau(250\,\mu{\rm})/A_{\rm J}=0.45\times
10^{-3}$ but dust opacity is known to be higher in molecular clouds
\citep{Planck2011b,Planck2014A-XI-allsky} and especially in dense
cores \citep{Stepnik2003,Roy2013}. Therefore, we tested cases where
the long-wavelength dust opacity was in dense regions increased to
$\tau(250\,\mu{\rm})/A_{\rm J}=1.6\times 10^{-3}$. This value was
derived from {\em Herschel} observations of cloud cores and could be
considered an upper limit for PGCC-type larger objects \citep{GCC-V}.

It would be tempting to interpret the difference between the simulated
and PGCC flux densities as proof of increased sub-millimetre opacity.
However, keeping the radiation field fixed, higher
sub-millimetre opacities did not translate into higher source fluxes.
When the long-wavelength opacity was increased, the dust temperatures
decreased. For the model $M_5$, the drop was about 2\,K
(Fig.~\ref{fig:plot_mod}d), almost compensating for the higher
opacity. The size of the extracted clumps in $M_5$ also decreased by
almost a factor of two (Fig.~\ref{fig:plot_mod}e). The alternative
density threshold $n({\rm H})=1000\,{\rm cm}^{-3}$ resulted in much
smaller changes. Although source fluxes might be increased by fine
tuning the density threshold, it seems unlikely that the flux
differences can be explained by dust properties. Conversely, our
simulations do not exclude the possibility of the PGCC clumps having
increased sub-millimetre opacity. One needs comparisons with other
column density tracers, such as near-infrared extinction, to get
direct constraints on the dust opacity \citep{Martin2011, Roy2013,
GCC-V}.

For models $L$ with longer LOS, the background intensities are
naturally higher. However, also the clump flux densities are strongly
correlated with the LOS length (Fig.~\ref{fig:plot_mod}a). This is due
to the higher column densities and partly due to the increased size of
the extracted clumps. In the case of $L=3-4$, the flux densities are
comparable to the PGCC values. With the Planck 353\,GHz optical depth
map at 1$\degr$ resolution and the assumption of dust opacity
$\tau(353\,{\rm GHz})=6\times 10^{-27}\,N_{\rm H}$
\citep{Planck2014A-XI-allsky}, we estimate that the median column
density at the PGCC source positions is $N({\rm H})=8.3\times
10^{21}\,{\rm cm}^{-2}$ ($N({\rm H})=6.9\times 10^{21}\,{\rm cm}^{-2}$
for the FLUX\_QUALITY=1 sub-sample). This is more than twice the
column density of our default model and thus in qualitative agreement
with the finding that models with higher column densities are in
better agreement with the PGCC observations.

If the PGCC clumps are real compact 3D objects, their observed
properties should ideally be independent of the LOS length. However,
we find the properties to be affected by other LOS emission. A longer
LOS increases the average signal, which results in larger structures
with higher integrated flux densities appearing above the noise. This
is not a major factor in our simulations where the effect of $L$ on
the clump size is less than 20\%. A longer LOS also means more
confusion noise that leaves many of the fainter sources below the
detection limit. Figures~\ref{fig:compare_clump_counts} indeed show
that as the LOS is increased from $L=1$ to $L=4$, the number of
sources decreases by one third. The confusion noise and the
resulting selection effects thus explain the $F_{\nu}$ vs. $L$
correlation.

Although the observed PGCC properties are affected by projection
effects, the detections algorithm itself makes use of the signature of
cold dust emission. This sets preference to objects with large optical
depths (also in directions perpendicular to the LOS) and thus with
large volume densities.
LOS confusion is likely to be a more important for continuum
catalogues where the detections are based only on source brightness,
without further physical constraints. 
The problem could be alleviated only by using radial velocity
information from line measurements to identify and even separate
objects along the LOS.  

Increased observational noise reduced the number of detections
(Fig.~\ref{fig:compare_clump_counts}) and, unlike longer LOS,
decreased the average size of the extracted clumps
(Fig.~\ref{fig:plot_mod}e). 
In tests involving discrete heating sources, effects were seen
mainly in diffuse regions (Fig.~\ref{fig:plot_star_Tmap}), with no
major consequences for the clump statistics. We emphasise that these
tests were concerned with variations in the radiation field external
to the clumps, not the effect of protostellar sources embedded in the
clumps. That would require simulations with higher spatial resolution
and more careful source modelling \citep{Malinen2011}. 

The separation of cold clumps from their warmer envelopes is not
straightforward and requires observations in a wide wavelength range,
preferably at a high spatial resolution. In real observations, this
also applies to the small-grain emission, which may have some effect
via its contribution to the 100\,$\mu$m band and could be traced with
additional shorter-wavelength bands. Similarly, the cold-clump
emission can be contaminated by embedded hot sources, which should be
quantified with further mid-infrared observations. Strongly heated
clumps should of course not be detected as cold sources but they may
affect the detectability and estimated properties of even nearby
clumps \citep{PGCC}.

\subsection{Clump polarisation} \label{dis:pol}

In the study of clump polarisation, grain alignment efficiency was set
constant, simply scaling the maximum theoretical polarisation fraction
to $p=$20\%. All variations observed in $p$ are therefore caused by the
geometry of the magnetic field. 

The mean magnetic field was parallel to the $y$ axis. When LOS was 
perpendicular to the mean field, $p$ values of the clumps ranged from
$p\sim 5$\% to over 10\%, brighter clumps having lower polarisation
fractions. For LOS parallel to the mean field, $p$ values were 3-4
times lower and uncorrelated with the clump intensity. The radial $p$
profiles were also qualitatively different. When LOS was perpendicular
to the mean field, the average $p$ profile decreased by $\Delta
p\sim$1\% towards the clump centre, over a distance of $\sim10\arcmin$
(Fig.~\ref{fig:profiles_p}). The drop was larger for larger cloud
distances where this angular scale corresponds to cloud rather than
core scales. When LOS was parallel to the field, the values increased
towards the centre but on average less than $\Delta p=1$\%. Although
line tangling usually leads to geometrical depolarisation, when the
field is initially parallel to the LOS, the twisting of magnetic field
can only increase the observed polarisation. The effect was clear in
simulations and should be detectable also in observations.
Appendix~\ref{sect:PGCC_angles} shows a tentative test of this but
full analysis of polarisation fraction variations in the PGCC clumps
will be presented in \citet{Ristorcelli2019}.

At large scales, we observe similar anticorrelation between column
density and polarisation fraction as discussed for example in
\cite{Planck_2015_XX} (see Appendix~\ref{app:p},
Fig.~\ref{fig:plot_p_vs_I_fullmap}).  Although the clump
polarisation fraction is also anti-correlated with column density, its
average value is not significantly different from the average over the
whole projected model, $p\sim 10$\% for the $x$ and $z$ view
directions (cf. Fig.~\ref{fig:profiles_p}). The values are also not
sensitive to the resolution of the observations
(Fig.~\ref{fig:plot_p_vs_I_fullmap}).

\begin{figure}
\includegraphics[width=8.8cm]{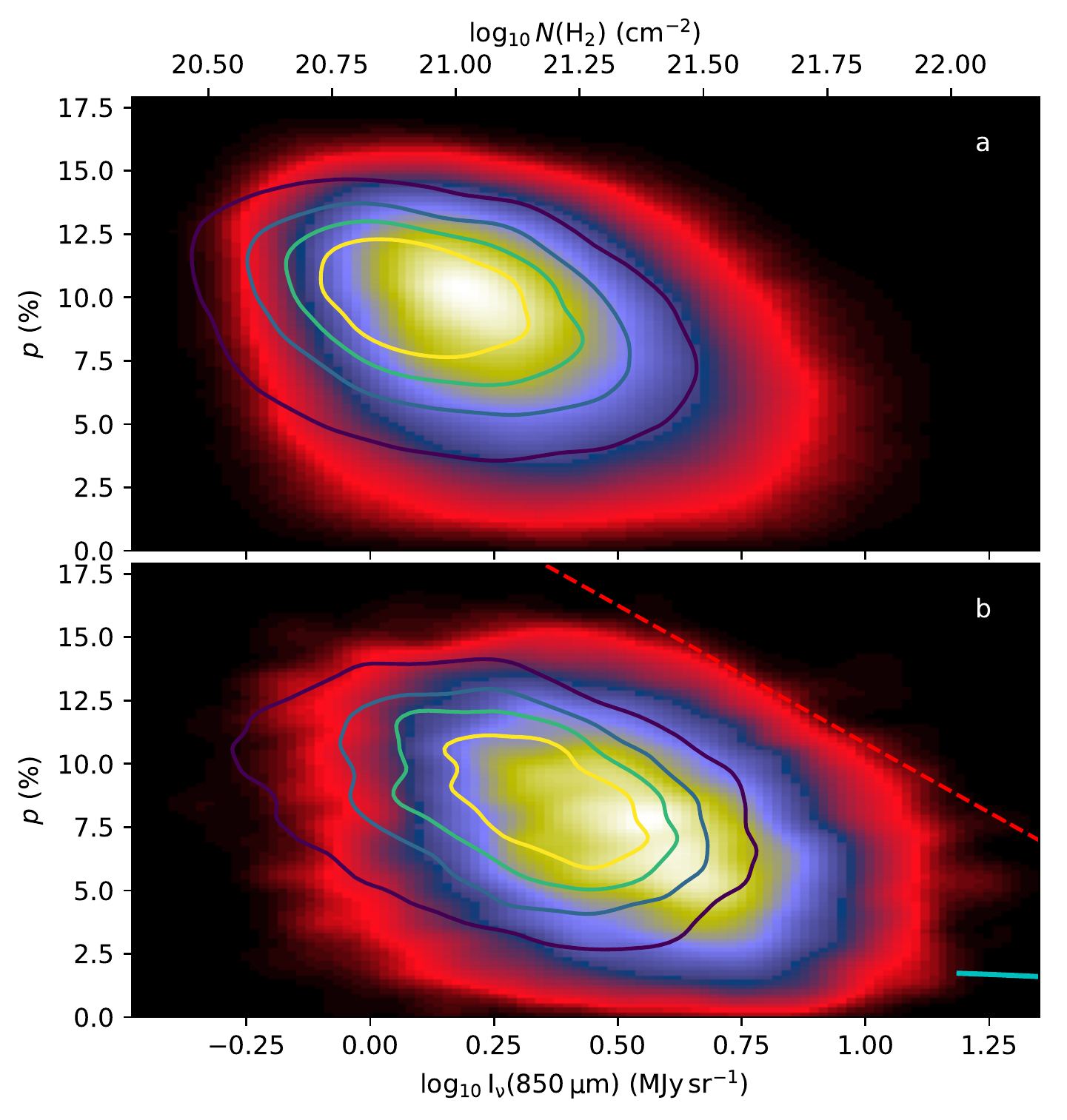}
\caption{ 
Relations $p$ vs. $I_{\nu}$ for clump centres (colour images) and for
background at 30$\arcmin$ distance (contours, in steps of 0.2 from 0.1
to 0.9 of the peak value). The plots include all clumps at 
$d=231$\,pc viewed from the $x$ and $y$ directions (frame a) and a
subset with ${\rm S/N}>10$ and $T_{\rm d}<13$\,K (frame b). The dashed red
line in frame b shows the upper envelope derived in
\citet{Planck_2015_XX}. The cyan line is the relation estimated for
the dense filament G035.39-00.33 \citet{Juvela2018_POL} and is drawn
for $N({\rm H}_2)>10^{22}\,{\rm cm}^{-2}$. The column density scale
corresponding to $T_{\rm d}=15$\,K.
}
\label{fig:plot_p_vs_I}
\end{figure}

Figure~\ref{fig:plot_mod_radial} showed the radial profiles of $p$ and
the general dependence between $p$ and surface brightness.
Figure~\ref{fig:plot_p_vs_I} shows the data in a different form,
looking at the $p$ vs. $I_{\nu}$ relations separately for the centre
of the clumps and for background points selected at 30$\arcmin$
distance. We also plot the upper-envelope relation $p_{\rm max}=-10.9
\log_{10}(N({\rm H}))+252.0$ that was \citet{Planck_2015_XX} used to
fit both Planck observations and simulated data. This matches also
the upper envelope of the $p$ vs. $I_{\nu}$ relation of our synthetic
clumps. For reference we also include the relation $p=2.02-0.28\times
10^{22}\,N({\rm H}_2)$ derived for the massive filament G035.39-00.33
\citep{Juvela2018_POL}. This was estimated for column densities
$N({\rm H}_2)>10^{22}\,{\rm cm}^{-2}$, values almost completely beyond
the range of column densities probed by our synthetic low-resolution
observations. 

The simulated polarisation fraction depends only on the field geometry
that was characterised with $\langle \gamma \rangle$, $S_{\rm LOS}$,
and $S_{\rm POS}$. Only $S_{\rm POS}$ is available for observers
because $\langle \gamma \rangle$ and $S_{\rm LOS}$ depend on field
geometry along the LOS. The observed $p$ was strongly correlated with
all three parameters. The correlation coefficient was largest between
$p$ and $S_{\rm LOS}$, some $r=-0.96$ in the case of the mixed sample
of different view directions (Fig.~\ref{fig:its}). Comparison with
$\langle \gamma \rangle $ gave a value of $r=-0.81$ and comparison
with $S_{\rm POS}$ still a very significant value of $r=-0.65$. This
is not surprising because $S_{\rm POS}$, $\langle \gamma \rangle$ are
naturally correlated: when $\gamma$ is close to $90\degr$ and the
magnetic field is mainly along the LOS, the angles of the B field
projected onto POS will have a large scatter. This applies both to the
behaviour along single LOS ($S_{\rm LOS}$) and the resulting
polarisation angle dispersion on the sky ($S_{\rm POS}$). However,
while the correlations between $S_{\rm LOS}$ and $\langle \gamma
\rangle$ on one hand and $S_{\rm LOS}$ and $S_{\rm POS}$ on the other
were both strong ($r>0.6$), the correlation between $S_{\rm POS}$ and
$\langle \gamma \rangle$ was significantly lower (Fig.~\ref{fig:its}).
It was boosted by the mixture of view directions and disappeared in
observations along the mean-field direction (Fig.~\ref{fig:its_y}). 

The longer LOS model (Sect.~\ref{sect:longer}) were associated with
some decrease in $S_{\rm POS}$. In \citet{Mangilli2019}, similar
effect was seen in {\it PILOT} balloon-borne telescope \citep{PILOT}
observations of low Galactic latitudes and interpreted as averaging
over many turbulent cells, which recovers the mean Galactic field
orientation. In our simulations, the effect is also strong because of
the relatively uniform mean-field direction.

The alternative models (Table~\ref{table:models}) introduced minor
changes in the polarisation relations. Increased dust opacity in dense
regions leads to a decrease in the $p$ values by $\Delta p \sim$1\%
(Fig.~\ref{fig:plot_mod_radial}). According to
Fig.~\ref{fig:plot_mod_pol}, this is caused mainly by an increased
dispersion $S_{\rm LOS}$. If this were due to the different relative
weighting of LOS areas, this would imply that clumps have a more
uniform magnetic field compared to the rest of the 250\,pc LOS.
However, there are also other factors, for example, the clump
selection. Figures~\ref{fig:its_xz}-~\ref{fig:its_y} show that the $y$
direction is associated with much lower $p$ and higher $S_{\rm LOS}$
values, which leads to the almost bimodal distributions of
Fig.~\ref{fig:plot_mod_pol}. Thus, the changes (lower $p$, higher
$S_{\rm LOS}$) can be explained by a larger fraction of $y$-direction
clumps. The alternative dust model $M_5$ does indeed increase the
relative amount of $y$-direction clumps by some $6$\%. When the effect
of modified dust is examined for each direction separately, the shapes
of the $p$ and $S_{\rm LOS}$ distributions do not change for the $y$
direction but for the $x$ and $z$ directions $p$ moves towards lower
values and $S_{\rm LOS}$ towards higher values. This is consistent
with increased dust opacity giving more weight for regions with more
tangled fields. The effects in Fig.~\ref{fig:plot_mod_pol} thus have
two causes, the larger fraction of $y$-direction clumps and the
enhanced geometrical depolarisation.

The only noticeable effect of discrete heating sources was in the
$p$ distribution where the peak at $p\sim 2\%$ is slightly stronger
and the number of clumps with $p>10\%$ is higher
(Fig.~\ref{fig:plot_mod_pol}). The relative number of clumps detected
towards the three view directions is not affected. The tail towards
high $p$ is thus probably related to the increased temperature
contrast, which results in the detection of more tenuous clumps with
higher polarisation fractions. The heating of diffuse material 
increases its contribution to the polarised intensity, also
contributing to the steepening of the radial $p$ profiles of the
clumps (Fig.~\ref{fig:plot_mod_radial}).

Figure~\ref{fig:plot_mod_radial} showed $p$ profiles also for the
cases of increased observational noise. We emphasise that this noise
applies only to the observations leading to the detection and basic
characterisation of the clumps. The polarisation data were simulated
without observational noise to avoid the complications biased $p$
estimates. Higher noise level should lead to the detection of clumps
with preferentially higher column densities and thus lower $p$.
However, in Fig.~\ref{fig:plot_mod_radial} noise has increased
the polarisation fractions at the clump centre by $\Delta p \sim$1\%
(Fig.~\ref{fig:plot_mod_radial}b). However, this again results from a
change in the relative number of clumps for the different view
directions. Increased noise reduced the relative number of the 
$y$-direction clumps from 35.5\% to 24.1\%. Because of the much lower
$p$ values of the $y$-direction clumps, this affects the overall
polarisation statistics.

Larger column densities provide more chances for geometrical
depolarisation and the net effect on $p$ was mostly negative also in our
simulations. 
In Fig.~\ref{fig:correlate_p_vs_ITS_LOS}, $p$ decreases with $L$,
especially on the side of high $p$ values.  Furthermore, the $p$
values tend to decrease more outside the clumps, making the radial
profiles flatter (Fig.~\ref{fig:plot_mod_radial}). At the clump
centre, the average polarisation fraction is reduced by $\Delta
p$=1\%. Relative to the default models, $p$ increases only at the
highest surface brightness values (Fig.~\ref{fig:plot_mod_radial}c),
in a small fraction of all clumps. However, the effect might be
detectable for the real PGCC clumps that, as discussed in
Sect.~\ref{sect:alt}, typically correspond to higher column densities.
The general anticorrelation between $p$ and $N({\rm H})$ is of course
well known based on previous observations and simulations
\citep{Vrba1976, Gerakines1995, WardThompson2000, planck2014-XIX,
Alves2014_pol, Planck_2015_XX_2, Planck2018_XII, Juvela2018_POL,
Seifried2019, Coude2019}.  

Our results on polarisation fraction are qualitatively similar to the
core-scale simulations of \citet{Chen2016_B}, where strong
correlations were also observed between $p$, $\langle \gamma \rangle$,
and $S_{\rm LOS}$. \citet{Chen2016_B} found that, compared to field
inclination effects, polarisation fraction decreases during later core
evolution relatively more because of the increased field tangling.
Compared to those simulations, our clumps represent both larger scales
and earlier pre-stellar stages. Another difference is the contribution
from the LOS, outside the main clump. Our 250\,pc LOS prevents the
observed $p$ values from becoming very low even towards the densest
clumps (view direction $y$ excluded). Based on the difference in
column densities, such LOS contributions should be even stronger in
the case of the low-latitude PGCC clumps. Another question related to
the LOS confusion is how the background affects observations from
interferometers and ground-based instruments, when the extended
emission is filtered out from the observations. The importance of this
on the recovered field morphology is, however, outside the scope of
this paper. 
%



The comparison of these simulations and observations, especially
those of the PGCC clumps, can be used to address further the questions
of dust opacity and grain alignment. Figure~\ref{fig:plot_mod_radial}a
showed that the surface brightness contrast between the clump centre
and the environment at 30$\arcmin$ distance is for the $T_{\rm
d}<13$\,K sample only $\sim$1.65. For the corresponding sample of PGCC
clumps this ratio is close to two \citep{Ristorcelli2019}. However, in
our tests the modified-dust-opacity models $M_1$ and $M_5$ did not
significantly change the contrast and it may be more sensitive to the
details of the clump selection than the dust properties. For example,
both noise and background fluctuations (LOS confusion) directly lead
to an increase in the average brightness of the detected clumps.

The variations of dust polarisation are more interesting.  The
polarisation fraction $p$ drops towards the simulated clumps on
average by less than $\Delta p=$2\%. For the $y$ direction, when LOS
is parallel to the mean field, the polarisation even increases towards
the clumps. For the $x$ and $z$ view directions, the average ratio of
$p$ values measured at the clump centre and in the background at
30$\arcmin$ distance is 0.85. For the PGCC, the corresponding factor is
$\sim0.60$. This is thus significantly smaller, even though the PGCC
sources should correspond to a mixture of different LOS vs. B-field
configurations. This suggests that the polarised emission from dense
clumps is reduced by additional factors, such as the RAT mechanism.

Assuming that RAT is the main grain alignment mechanism, polarised
emission should be reduced in regions with weaker radiation fields. In
the PGCC and in our simulations, the sources are selected directly based
on their spectral cold-dust signature, which means that their inner
regions are strongly shielded from the interstellar radiation field.
Because grain alignment is disturbed by collision, the other main
parameter is the local density. Early numerical studies predicted
clear drop of emitted polarised intensity from clumps shielded by
$A_{\rm V}$ of a few and densities in MHD simulations above $n({\rm
H_2})\sim 10^{3}$\,cm$^{-3}$ \citep{Bethell2007, Pelkonen2007,
Pelkonen2009}. With more recent molecular-cloud simulations,
\cite{Seifried2019} concluded that RAT would remain effective up to
densities $n({\rm H_2})\sim 10^{4}$\,cm$^{-3}$. Observations do show a
strong drop in $p$ towards dense filaments and cores
\citep{Brauer2016, Juvela2018_POL, Kandori2018} but in individual
sources it is difficult to disentangle the effects of dust physics
from those of the field geometry. According to RAT theory the
polarisation fraction should drop continuously with increasing volume
density. The actual effect thus depends critically on the nature of
the studied objects. The PGCC contains a very heterogeneous collection of
sources, from nearby cloud cores (average densities above
$10^{4}$\,cm$^{-3}$) to more distant extended cloud structures,
sometimes with very high column densities (such as infrared dark
clouds). Large uncertainties are still associated with the dust
properties, what is the grain size distribution inside the cold clumps
and how grain evolution affects the emission and the alignment
\citep{Pelkonen2009,Reissl2017}. The polarisation fraction of the PGCC
clumps will be investigated further in \citet{Ristorcelli2019}. This
analysis and further modelling of Planck polarisation data should
shed more light on these questions.

\section{Conclusions} \label{sect:conclusions}

We have used numerical simulations of interstellar clouds to make
synthetic observations of a large number of cold clumps. The clumps
were extracted with a detection method similar to that used for the
PGCC \citep{PGCC}. We compared the properties of the simulated clumps
with those of the PGCC sources. We also examined the polarisation
fraction variations associated to the clumps. The study has led to the
following conclusions.

Many physical clump properties (e.g. sizes, aspect rations, and
temperatures) of the simulated clumps are very similar to those in the
Planck survey. However, especially the clump size distributions are
determined mainly by the angular resolution of the observations and
the upper limit set by the detection method. 

The column densities of the synthetic clumps are lower than in the PGCC.
This was concluded to be caused mainly by the model column densities
that are lower than those encountered in the PGCC at low and intermediate
Galactic latitudes. The observed clump column density distribution can
be matched by increasing the average column density by a factor of
2-3. This suggests that the derived clump properties, also in the PGCC,
are not properties of well-defined compact 3D sources but do depend on
LOS confusion. The apparent difference in column densities can not be
interpreted as direct evidence of increased dust sub-millimetre
emissivity, although the simulations cannot exclude that possibility
either.

The clumps are usually associated with a small decrease in the
polarisation fraction, $\Delta p \sim 1$\%, relative to their
surroundings. In the more rare case, where the mean magnetic field
parallel to the LOS, $p$ tends to increase towards the clumps. This is
caused by the very low background values and is typically less than
$\Delta p=0.5$\%.

The alternative models show some noticeable changes in the
polarisation fraction. Higher dust opacity (model $M_5$) and increased
LOS (model $L_3$) decrease the $p$ values towards clump centres by
$\Delta p=1$\%. In the case of the latter, polarisation remains lower
also outside the clumps and thus results in a low contrast in the $p$
values between the clumps and their environment. Larger
observational noise increases the average column density and decreases
the average $p$ of the extracted clumps. On the other hand, the
discrete heating sources, used to increase the level and variations of
the radiation field within the model volume, have but a marginal
effect on the polarisation and other clump properties. 

The surface brightness contrast between the clumps and their
background is smaller for the simulated clumps than for the PGCC
clumps.  The drop in $p$ is also smaller in the simulations, which
suggests that the polarised emission from the PGCC clumps may be affected
by additional factors, such as the imperfect grain alignment predicted
by the RAT mechanism. This will be addressed in future studies.

\begin{acknowledgements}
MJ acknowledges the support of the Academy of Finland Grant No.
285769.
PP acknowledges support by the Spanish MINECO under project
AYA2017-88754-P.
VMP acknowledges support by the Spanish MINECO under projects
MDM-2014-0369 and AYA2017-88754-P.
We thankfully acknowledge the computer resources at MareNostrum and
the technical support provided by Barcelona Supercomputing Center
(AECT-2018-3-0019).
\end{acknowledgements}

\bibliography{MJ_bib}

\begin{appendix}

\section{Additional plots of polarisation quantities} \label{app:p}

\begin{figure} 
\includegraphics[width=8.8cm]{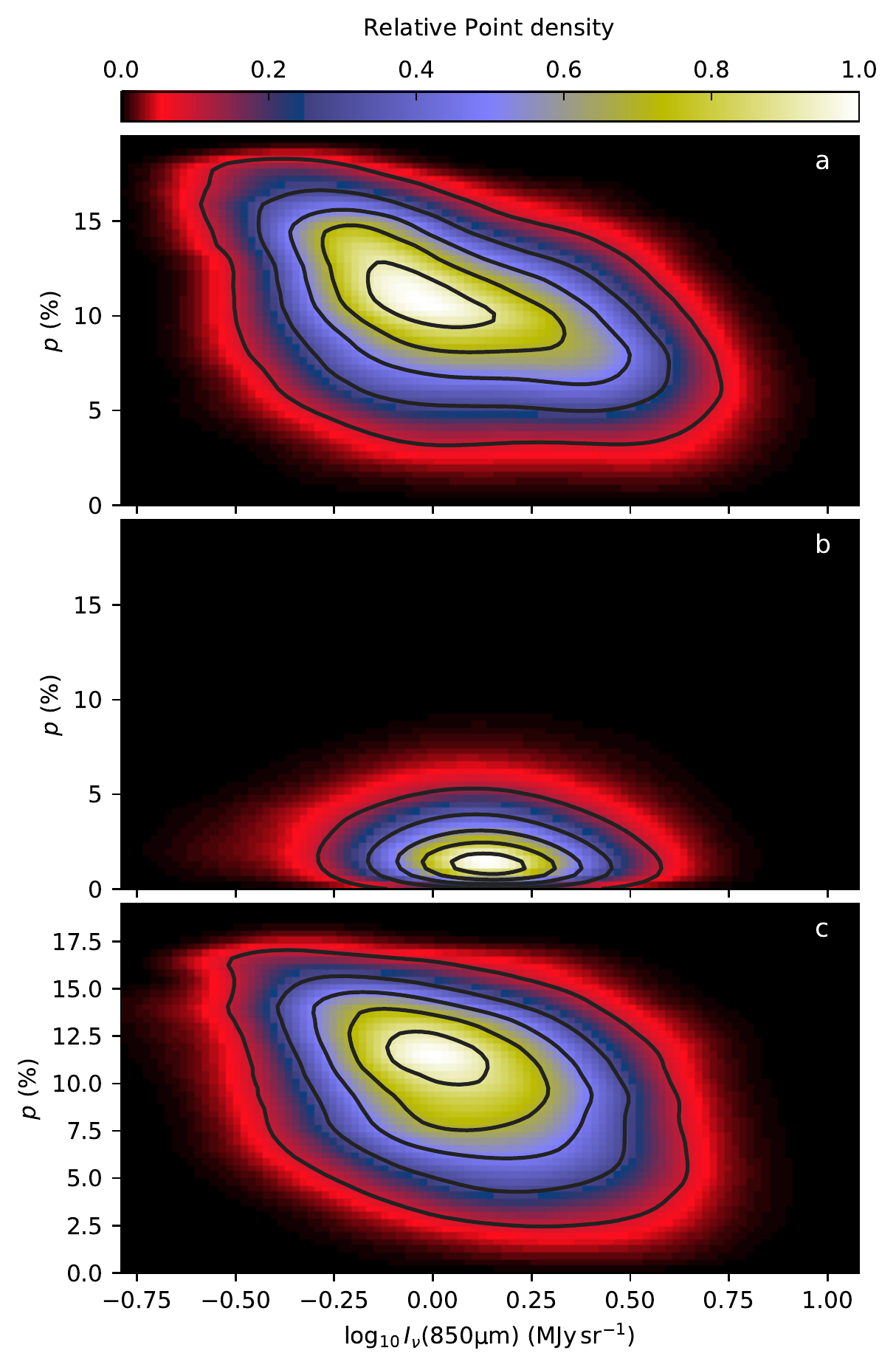}
\caption{
Polarisation fraction as function of 850\,$\mu$m intensity over whole
snapshot 377 (model $D$). The frames a-c correspond, respectively, to
the view directions $x$, $y$, and $z$. The colour images show the
distribution for the full-resolution maps and the contours (drawn in
steps of 10\% between 10\% and 90\% of the maximum value) for
lower-resolution data that correspond to a distance of 351\,pc.
}
\label{fig:plot_p_vs_I_fullmap}
\end{figure}

\begin{figure*}  
\sidecaption
\includegraphics[width=12cm]{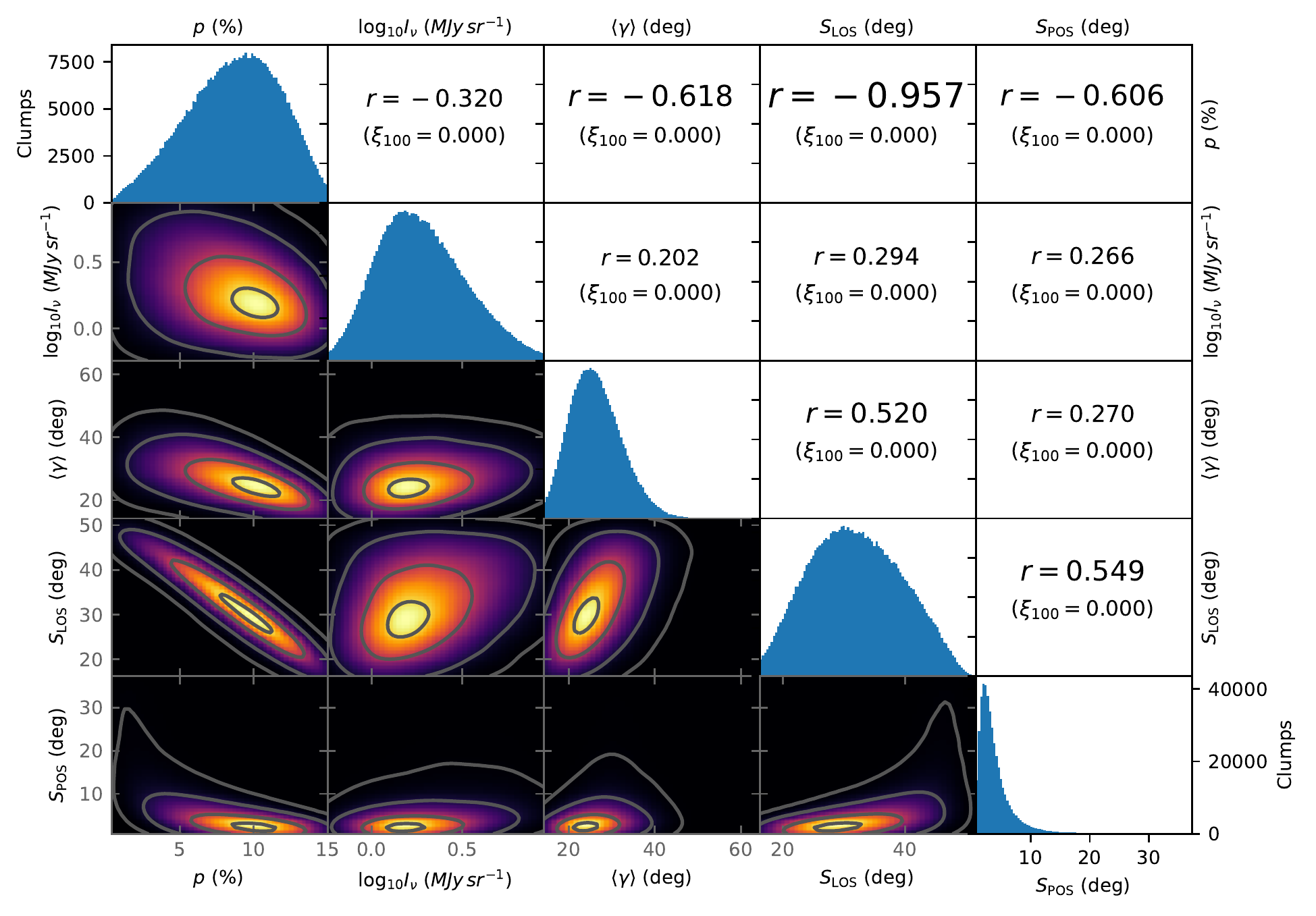}
\caption{
Same as Fig.~\ref{fig:its} but only for clumps viewed from  $x$ and
$z$ directions. Upper right frames include probabilities $\xi_{100}$
for $r$ to be consistent with zero. These are calculated using random
samples that are a factor of 100 smaller than the full clump samples.
}
\label{fig:its_xz}
\end{figure*}

\begin{figure*} 
\sidecaption
\includegraphics[width=12cm]{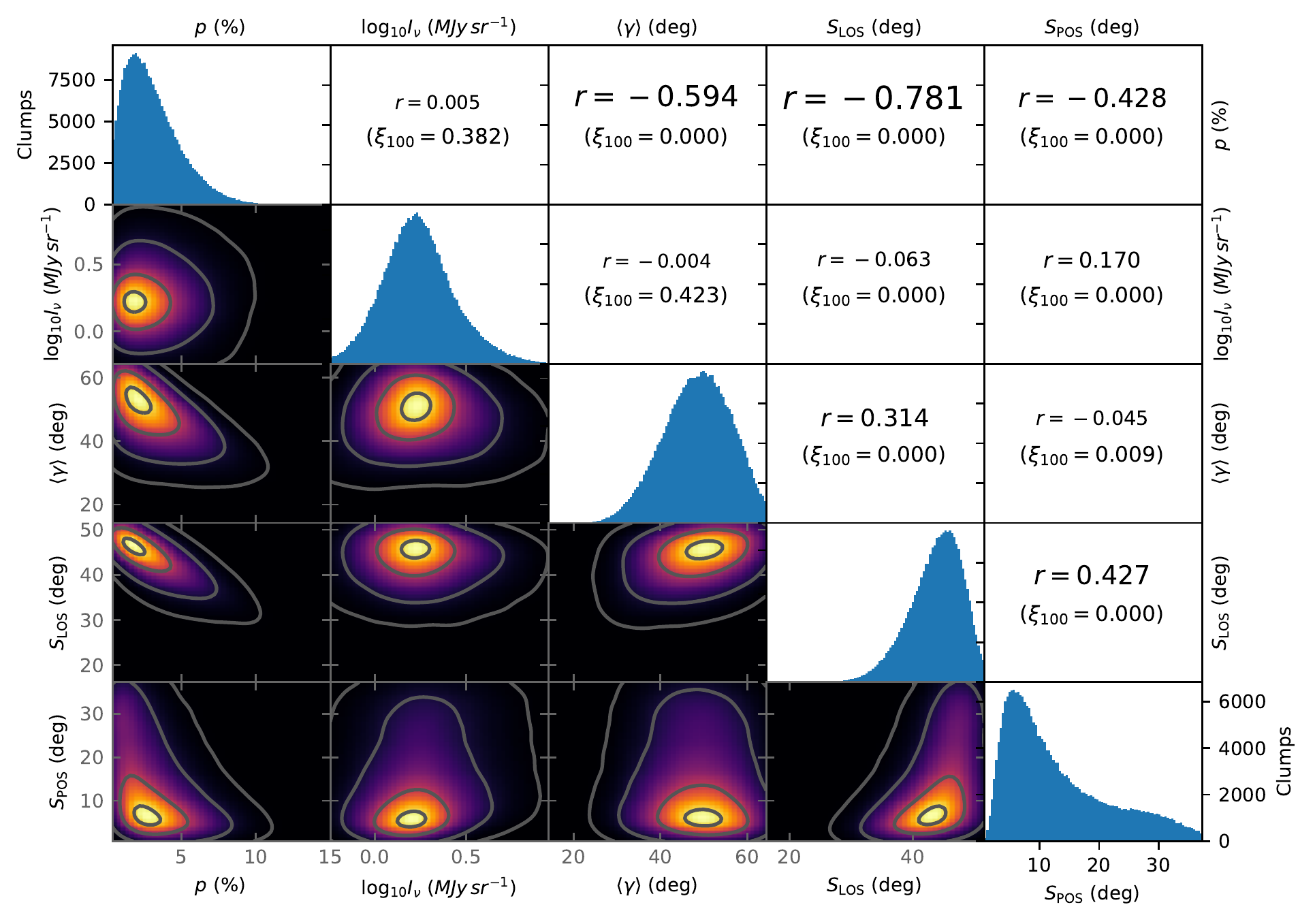}
\caption{
Same as Fig.~\ref{fig:its} but only for clumps viewed from $y$
direction.
}
\label{fig:its_y}
\end{figure*}

\begin{figure*}
\sidecaption
\includegraphics[width=12cm]{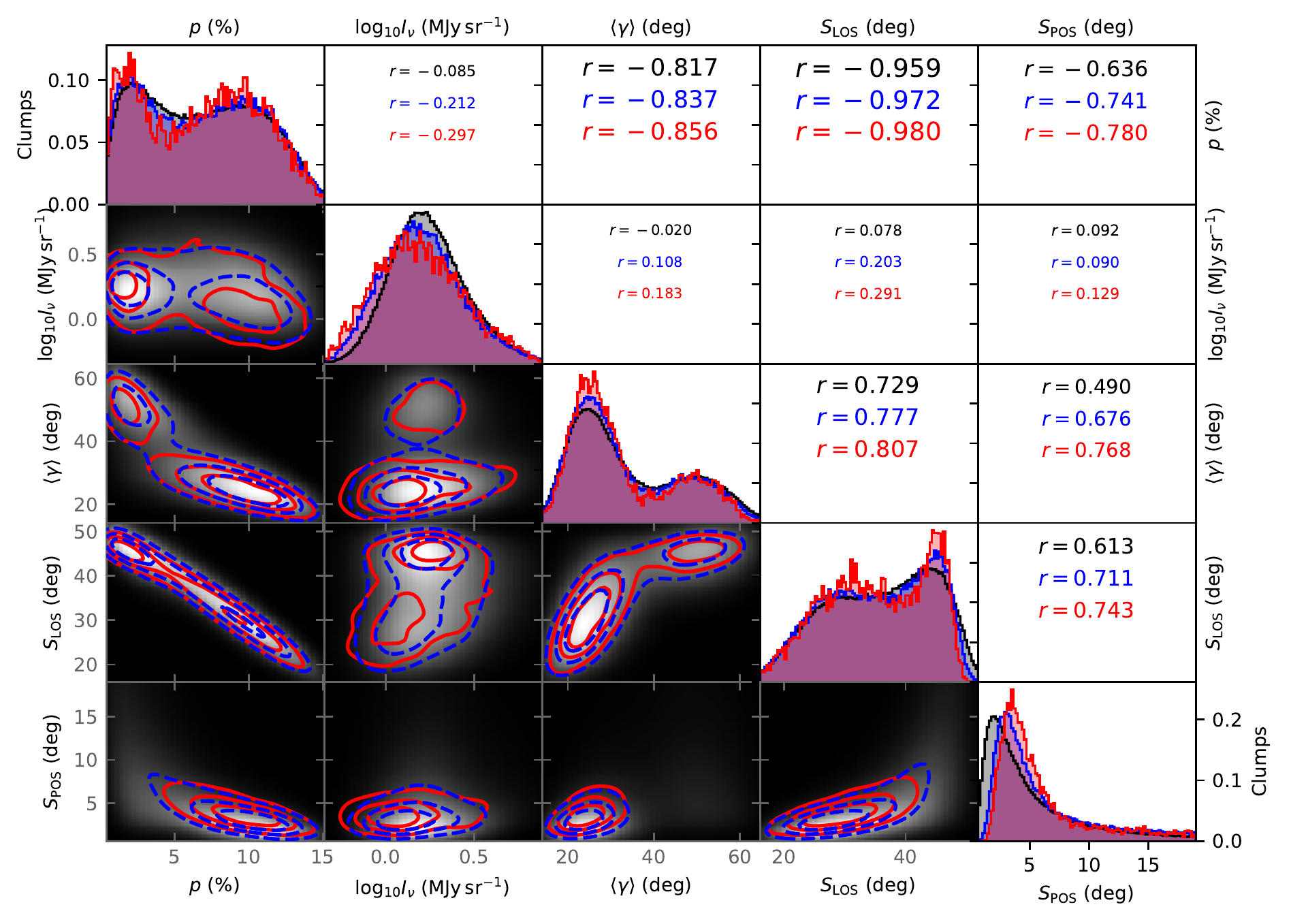}
\caption{
As Fig.~\ref{fig:its} but comparing default models at different
distances. The background images and histograms plotted with black
lines are for $d$=152\,pc. Data for distances $d$=533\,pc and
$d=1232$\,pc are drawn with blue and red lines (and contours), 
respectively. The correlation coefficients in the frames above the
diagonal are written in the corresponding colours.
}
\label{fig:plot_pcor_distance}
\end{figure*}

Figures~\ref{fig:its_xz}-~\ref{fig:its_y} show the correlations of the
clump polarisation fraction $p$ with other parameters. The figures are
similar to Fig.~\ref{fig:its} but show the data separately for the
combination of $x$ and $z$ directions where the LOS is mainly
perpendicular to the mean field direction (Fig.~\ref{fig:its_xz}) and
for the $y$ direction where it is mainly parallel to the field
(Fig.~\ref{fig:its_y}). 
Figure~\ref{fig:plot_pcor_distance} compares the same parameters for
the default models at three distances, 152\,pc, 533\,pc, and 1232\,pc.

\section{Polarisation fraction of PGCC clumps} \label{sect:PGCC_angles}

The full analysis of polarisation fraction variations will be
presented in \citet{Ristorcelli2019}. However, we make here a
preliminary test to see if the dichotomy of $p$ profiles is detectable
also for the real PGCC clumps (cf Sect.~\ref{dis:pol}). When LOS is
perpendicular to the mean field direction, simulations showed several
percent higher polarisation fractions, $p$ decreasing towards the
clump centre. When LOS was parallel to the mean field, the clumps had
higher polarisation than their environment. However, these
radial variations were on average less than 1\%.

\begin{figure}
\sidecaption
\includegraphics[width=8.8cm]{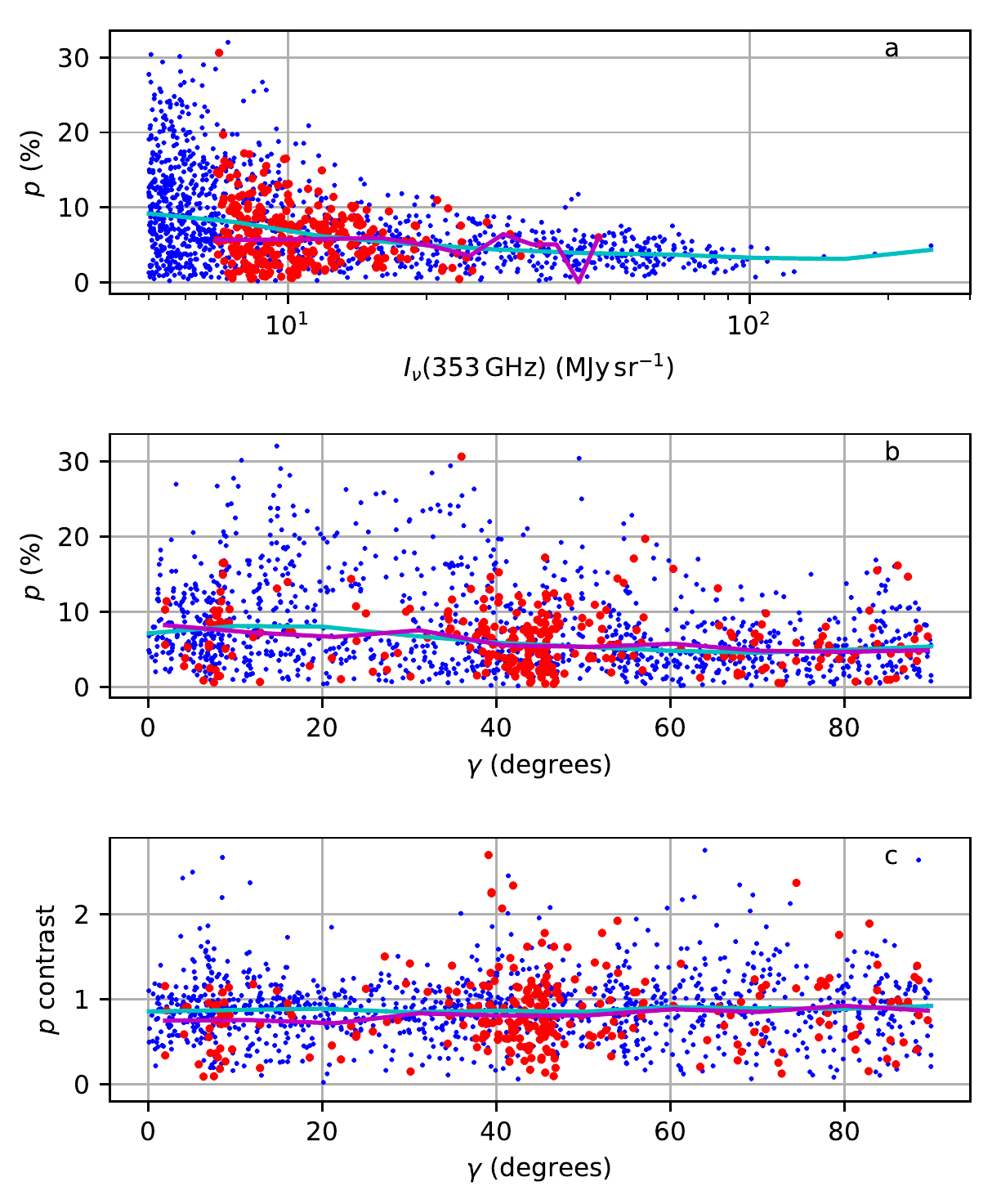}
\caption{
Polarisation fraction associated to selected PGCC clumps. The first
frame shows the relation $p$ vs. intensity when both are measured at
the clump centres. The second frame shows $p$ as a function of the
estimated $\gamma$ angle. The last frame shows the same for the $p$
contrast ($p$ at the clump centre divided by the average over radial
distances 16-20$\arcmin$). The blue and red colours correspond to two
clump samples (see text). The cyan and magenta lines show the
corresponding median values calculated with a moving window with the
width one tenth of the data range.
}
\label{fig:PGCC_p_vs_angles}
\end{figure}

We calculated $p_{\rm max}$ estimates from the Planck all-sky maps
and extracted radial intensity and $p$ profiles for the PGCC clumps.
This was done in a similar way as in the analysis of the simulated
data (cf. Sect.~\ref{sect:p1}). However, to decrease the effects of
noise (present in observations but not in our simulations), the input
maps were first convolved to 10$\arcmin$ resolution.
Figure~\ref{fig:PGCC_p_vs_angles} shows $p$ as function of intensity
and the estimated $\gamma$ angle. This angle is calculated based on
the Galactic coordinates and the distance of the PGCC clumps, assuming
that Milky Way would have an ordered large-scale field. We assume that
the field would be main azimuthal, with a pitch angle of 14$\degr$, as
estimated for the spiral structure \citep{Vallee2017AstRv}. The results are shown
in Fig.~\ref{fig:PGCC_p_vs_angles}.

The analysis is naturally limited to clumps with distance estimates.
We further limit the sample to sources with distances $d<8.5$\,kpc and
with central intensity higher than $I_{\nu}(353\,{\rm GHz})=5\,{\rm
MJy\,sr^{-1}}$. These sources are plotted in
Fig.~\ref{fig:PGCC_p_vs_angles} in blue colour. The scatter of $p$
values in increasing towards lower intensities. This is a sign of the
estimates are becoming more uncertain and possibly biased because of
the noise. We include in the plots also a more restricted set of
sources. This includes only clumps with intensities above $5\,{\rm
MJy\,sr^{-1}}$, with distances 200-5000\,pc, and Galactic latitudes
$b=2-30\degr$. The constraint were included in an attempt to eliminate
the most confused LOS in the Galactic plane and to avoid giving a high
weight for the nearby clouds that may have systematic deviations from
our very simple model for the $\gamma$ angles. This second set of
sources is plotted in red colour.

Figure~\ref{fig:PGCC_p_vs_angles}b is thus expected to show a
decrease in $p$ values between the perpendicular ($\gamma=0\degr$) and
the parallel ($\gamma=90\degr$) cases. Such as negative trend also is
observed, $p$ decreasing by almost $\Delta p=3$\% as $\gamma$ increases from
0$\degr$ to $90\degr$. The trend is similar for both clump samples.
The trend should in principle be statistically significant because
$\gamma$ is not expected to be strongly correlated with clump
intensity, which in turn is strongly correlated with the $p$
uncertainties and $p$ bias.

The angle between LOS and magnetic field has in simulations a smaller
effect on the shape of the radial $p$ profiles. In 
Fig.~\ref{fig:profiles_p}, the relative change between the clump
centre and values at distances $R>10\arcmin$ is only of the order of
10\%. Figure~\ref{fig:PGCC_p_vs_angles}c plots the contrast of $p$
values as a function the angle $\gamma$ for the PGCC clumps. The
contrast is here defined as $p$ in the clump centre divided by the
mean value over distances $R=16-20\arcmin$. There is a marginal trend
that agrees with the simulations: the contrast is smaller for small
$\gamma$ angles. However, the median values remain below one, even
when the field is supposed to be parallel to the LOS
($\gamma=90\degr$).

\end{appendix}

\end{document}